\documentstyle[epsf,epsfig,here,12pt]{article}  
\begin{document}

\newcommand{\bptre}{\rm b^{+}_{3}}
\newcommand{\bp}{\rm b^{+}_{1}}
\newcommand{\bo}{\rm b^0}
\newcommand{\bos}{\rm b^0_s}
\newcommand{\bss}{\rm b^s_s}
\newcommand{\qq}{\rm q \overline{q}}
\newcommand{\cc}{\rm c \overline{c}}
\newcommand{\BsDmX}{{B_{s}^{0}} \rightarrow D \mu X}
\newcommand{\BsDsm}{{B_{s}^{0}} \rightarrow D_{s} \mu X}
\newcommand{\BsDsX}{{B_{s}^{0}} \rightarrow D_{s} X}
\newcommand{\BDsX}{B \rightarrow D_{s} X}
\newcommand{\BDomX}{B \rightarrow D^{0} \mu X}
\newcommand{\BDpmX}{B \rightarrow D^{+} \mu X}
\newcommand{\Dsfmn}{D_{s} \rightarrow \phi \mu \nu}
\newcommand{\Dsfipi}{D_{s} \rightarrow \phi \pi}
\newcommand{\DsfX}{D_{s} \rightarrow \phi X}
\newcommand{\DpfX}{D^{+} \rightarrow \phi X}
\newcommand{\DofX}{D^{0} \rightarrow \phi X}
\newcommand{\DfX}{D \rightarrow \phi X}
\newcommand{\DsD}{B \rightarrow D_{s} D}
\newcommand{\DsmX}{D_{s} \rightarrow \mu X}
\newcommand{\DmX}{D \rightarrow \mu X}
\newcommand{\Zbb}{Z^{0} \rightarrow \rm b \overline{b}}
\newcommand{\Zcc}{Z^{0} \rightarrow \rm c \overline{c}}
\newcommand{\Rbb}{\frac{\Gamma_{Z^0 \rightarrow \rm b \overline{b}}}
{\Gamma_{Z^0 \rightarrow Hadrons}}}
\newcommand{\Rcc}{\frac{\Gamma_{Z^0 \rightarrow \rm c \overline{c}}}
{\Gamma_{Z^0 \rightarrow Hadrons}}}
\newcommand{\bb}{\rm b \overline{b}}
\newcommand{\str}{\rm s \overline{s}}
\newcommand{\Bs}{\rm{B^0_s}}
\newcommand{\Bsb}{\overline{\rm{B^0_s}}}
\newcommand{\Bp}{\rm{B^{+}}}
\newcommand{\Bm}{\rm{B^{-}}}
\newcommand{\Bo}{\rm{B^{0}}}
\newcommand{\Bd}{\rm{B^{0}_{d}}}
\newcommand{\Bdb}{\overline{\rm{B^{0}_{d}}}}
\newcommand{\Lb}{\Lambda^0_b}
\newcommand{\Lbb}{\overline{\Lambda^0_b}}
\newcommand{\Kstar}{\rm{K^{\star 0}}}
\newcommand{\phim}{\rm{\phi}}
\newcommand{\Ds}{\mbox{D}_s}
\newcommand{\Dsp}{\mbox{D}_s^+}
\newcommand{\Dp}{\mbox{D}^+}
\newcommand{\Dn}{\mbox{D}^0}
\newcommand{\Dsb}{\overline{\mbox{D}_s}}
\newcommand{\Dm}{\mbox{D}^-}
\newcommand{\Dnb}{\overline{\mbox{D}^0}}
\newcommand{\Lc}{\Lambda_c}
\newcommand{\Lcb}{\overline{\Lambda_c}}
\newcommand{\Dstarp}{\mbox{D}^{\ast +}}
\newcommand{\Dstarm}{\mbox{D}^{\ast -}}
\newcommand{\Dsstarp}{\mbox{D}_s^{\ast +}}
\newcommand{\Pb}{P_{b-baryon}}
\newcommand{\KKpi}{\rm{ K K \pi }}
\newcommand{\GeV}{\rm{GeV}}
\newcommand{\MeV}{\rm{MeV}}
\newcommand{\nb}{\rm{nb}}
\newcommand{\Zzero}{{\rm Z}^0}
\newcommand{\MZ}{\rm{M_Z}}
\newcommand{\MW}{\rm{M_W}}
\newcommand{\GF}{\rm{G_F}}
\newcommand{\Gm}{\rm{G_{\mu}}}
\newcommand{\MH}{\rm{M_H}}
\newcommand{\MT}{\rm{m_{top}}}
\newcommand{\GZ}{\Gamma_{\rm Z}}
\newcommand{\Afb}{\rm{A_{FB}}}
\newcommand{\Afbs}{\rm{A_{FB}^{s}}}
\newcommand{\sigmaf}{\sigma_{\rm{F}}}
\newcommand{\sigmab}{\sigma_{\rm{B}}}
\newcommand{\NF}{\rm{N_{F}}}
\newcommand{\NB}{\rm{N_{B}}}
\newcommand{\Nnu}{\rm{N_{\nu}}}
\newcommand{\RZ}{\rm{R_Z}}
\newcommand{\rhob}{\rho_{eff}}
\newcommand{\Gammanz}{\rm{\Gamma_{Z}^{new}}}
\newcommand{\Gammani}{\rm{\Gamma_{inv}^{new}}}
\newcommand{\Gammasz}{\rm{\Gamma_{Z}^{SM}}}
\newcommand{\Gammasi}{\rm{\Gamma_{inv}^{SM}}}
\newcommand{\Gammaxz}{\rm{\Gamma_{Z}^{exp}}}
\newcommand{\Gammaxi}{\rm{\Gamma_{inv}^{exp}}}
\newcommand{\rhoZ}{\rho_{\rm Z}}
\newcommand{\thw}{\theta_{\rm W}}
\newcommand{\swsq}{\sin^2\!\thw}
\newcommand{\swsqmsb}{\sin^2\!\theta_{\rm W}^{\overline{\rm MS}}}
\newcommand{\swsqbar}{\sin^2\!\overline{\theta}_{\rm W}}
\newcommand{\cwsqbar}{\cos^2\!\overline{\theta}_{\rm W}}
\newcommand{\swsqb}{\sin^2\!\theta^{eff}_{\rm W}}
\newcommand{\ee}{{e^+e^-}}
\newcommand{\eeX}{{e^+e^-X}}
\newcommand{\gaga}{{\gamma\gamma}}
\newcommand{\mumu}{\ifmmode {\mu^+\mu^-} \else ${\mu^+\mu^-} $ \fi}
\newcommand{\eeg}{{e^+e^-\gamma}}
\newcommand{\mumug}{{\mu^+\mu^-\gamma}}
\newcommand{\tautau}{{\tau^+\tau^-}}
\newcommand{\qqb}{{q\bar{q}}}
\newcommand{\eegg}{e^+e^-\rightarrow \gamma\gamma}
\newcommand{\eeggg}{e^+e^-\rightarrow \gamma\gamma\gamma}
\newcommand{\eeee}{e^+e^-\rightarrow e^+e^-}
\newcommand{\eeeeee}{e^+e^-\rightarrow e^+e^-e^+e^-}
\newcommand{\eeeeg}{e^+e^-\rightarrow e^+e^-(\gamma)}
\newcommand{\eeeegg}{e^+e^-\rightarrow e^+e^-\gamma\gamma}
\newcommand{\eeeg}{e^+e^-\rightarrow (e^+)e^-\gamma}
\newcommand{\eemumu}{e^+e^-\rightarrow \mu^+\mu^-}
\newcommand{\eetautau}{e^+e^-\rightarrow \tau^+\tau^-}
\newcommand{\eehad}{e^+e^-\rightarrow {\rm hadrons}}
\newcommand{\eettg}{e^+e^-\rightarrow \tau^+\tau^-\gamma}
\newcommand{\eell}{e^+e^-\rightarrow l^+l^-}
\newcommand{\Ztopig}{{\rm Z}^0\rightarrow \pi^0\gamma}
\newcommand{\Ztogg}{{\rm Z}^0\rightarrow \gamma\gamma}
\newcommand{\Ztoee}{{\rm Z}^0\rightarrow e^+e^-}
\newcommand{\Ztoggg}{{\rm Z}^0\rightarrow \gamma\gamma\gamma}
\newcommand{\Ztomumu}{{\rm Z}^0\rightarrow \mu^+\mu^-}
\newcommand{\Ztotautau}{{\rm Z}^0\rightarrow \tau^+\tau^-}
\newcommand{\Ztoll}{{\rm Z}^0\rightarrow l^+l^-}
\newcommand{\Ztocc}{{\rm Z^0\rightarrow c \bar c}}
\newcommand{\Lamp}{\Lambda_{+}}
\newcommand{\Lamm}{\Lambda_{-}}
\newcommand{\Pt}{\rm P_{t}}
\newcommand{\Gee}{\Gamma_{ee}}
\newcommand{\Gpig}{\Gamma_{\pi^0\gamma}}
\newcommand{\Ggg}{\Gamma_{\gamma\gamma}}
\newcommand{\Gggg}{\Gamma_{\gamma\gamma\gamma}}
\newcommand{\Gmumu}{\Gamma_{\mu\mu}}
\newcommand{\Gtautau}{\Gamma_{\tau\tau}}
\newcommand{\Ginv}{\Gamma_{\rm inv}}
\newcommand{\Ghad}{\Gamma_{\rm had}}
\newcommand{\Gnu}{\Gamma_{\nu}}
\newcommand{\GnuSM}{\Gamma_{\nu}^{\rm SM}}
\newcommand{\Gll}{\Gamma_{l^+l^-}}
\newcommand{\Gff}{\Gamma_{f\overline{f}}}
\newcommand{\Gtot}{\Gamma_{\rm tot}}
\newcommand{\Rb}{\mbox{R}_b}
\newcommand{\Rc}{\mbox{R}_c}
\newcommand{\al}{a_l}
\newcommand{\vl}{v_l}
\newcommand{\af}{a_f}
\newcommand{\vf}{v_f}
\newcommand{\ael}{a_e}
\newcommand{\ve}{v_e}
\newcommand{\amu}{a_\mu}
\newcommand{\vmu}{v_\mu}
\newcommand{\atau}{a_\tau}
\newcommand{\vtau}{v_\tau}
\newcommand{\ahatl}{\hat{a}_l}
\newcommand{\vhatl}{\hat{v}_l}
\newcommand{\ahate}{\hat{a}_e}
\newcommand{\vhate}{\hat{v}_e}
\newcommand{\ahatmu}{\hat{a}_\mu}
\newcommand{\vhatmu}{\hat{v}_\mu}
\newcommand{\ahattau}{\hat{a}_\tau}
\newcommand{\vhattau}{\hat{v}_\tau}
\newcommand{\vtildel}{\tilde{\rm v}_l}
\newcommand{\avsq}{\ahatl^2\vhatl^2}
\newcommand{\Ahatl}{\hat{A}_l}
\newcommand{\Vhatl}{\hat{V}_l}
\newcommand{\Afer}{A_f}
\newcommand{\Ael}{A_e}
\newcommand{\Aferb}{\bar{A_f}}
\newcommand{\Aelb}{\bar{A_e}}
\newcommand{\AVsq}{\Ahatl^2\Vhatl^2}
\newcommand{\Iwk}{I_{3l}}
\newcommand{\Qch}{|Q_{l}|}
\newcommand{\roots}{\sqrt{s}}
\newcommand{\pT}{p_{\rm T}}
\newcommand{\mt}{m_t}
\newcommand{\Rechi}{{\rm Re} \left\{ \chi (s) \right\}}
\newcommand{\up}{^}
\newcommand{\abscosthe}{|cos\theta|}
\newcommand{\dsum}{\Sigma |d_\circ|}
\newcommand{\zsum}{\Sigma z_\circ}
\newcommand{\sint}{\mbox{$\sin\theta$}}
\newcommand{\cost}{\mbox{$\cos\theta$}}
\newcommand{\mcost}{|\cos\theta|}
\newcommand{\epair}{\mbox{$e^{+}e^{-}$}}
\newcommand{\mupair}{\mbox{$\mu^{+}\mu^{-}$}}
\newcommand{\taupair}{\mbox{$\tau^{+}\tau^{-}$}}
\newcommand{\gamgam}{\mbox{$e^{+}e^{-}\rightarrow e^{+}e^{-}\mu^{+}\mu^{-}$}}
\newcommand{\fullskip}{\vskip 16cm}
\newcommand{\halfskip}{\vskip  8cm}
\newcommand{\quarskip}{\vskip  6cm}
\newcommand{\abitskip}{\vskip 0.5cm}
\newcommand{\ba}{\begin{array}}
\newcommand{\ea}{\end{array}}
\newcommand{\bc}{\begin{center}}
\newcommand{\ec}{\end{center}}
\newcommand{\be}{\begin{eqnarray}}
\newcommand{\eeq}{\end{eqnarray}}
\newcommand{\bes}{\begin{eqnarray*}}
\newcommand{\ees}{\end{eqnarray*}}
\newcommand{\Kz}{\ifmmode {\rm K^0_s} \else ${\rm K^0_s} $ \fi}
\newcommand{\Zz}{\ifmmode {\rm Z^0} \else ${\rm Z^0 } $ \fi}
\newcommand{\qqbar}{\ifmmode {\rm q\bar{q}} \else ${\rm q\bar{q}} $ \fi}
\newcommand{\ccbar}{\ifmmode {\rm c\bar{c}} \else ${\rm c\bar{c}} $ \fi}
\newcommand{\bbbar}{\ifmmode {\rm b\bar{b}} \else ${\rm b\bar{b}} $ \fi}
\newcommand{\xxbar}{\ifmmode {\rm x\bar{x}} \else ${\rm x\bar{x}} $ \fi}
\newcommand{\rphi}{\ifmmode {\rm R\phi} \else ${\rm R\phi} $ \fi}
%\renewcommand {\pt}         {\rm p_t}
%========================================================================% 

%################################################## titlepage declaration

\begin{titlepage}

\pagenumbering{arabic}
\vspace*{-1.5cm}
%\begin{tabular*}{15.cm}{l@{\extracolsep{\fill}}r}
%{  } & 
%%===================> DELPHI note number       =====> To be filled <=====%
%%CERN-PPE 97-137  
%DELPHI 97-137 PHYS 724 
%%========================================================================%
%\\
%& 
%%===================> DELPHI note date         =====> To be filled <=====%
%13 August, 1997
%%========================================================================%
%\\
%&\\ \hline
%\end{tabular*}
\vspace*{2.cm}
\begin{center}
\Large 
{\bf  Measurements of the $\rho$ and $\eta$ parameters of the $V_{CKM}$ matrix and perspectives. } 
\vspace*{2.cm}
\\
\normalsize { 
   {\bf 
  P. Paganini$^{*}$ , F. Parodi$^{+}$ , P. Roudeau and A. Stocchi }\\
\vskip 0.5truecm
   {\footnotesize Laboratoire de l'Acc\'el\'erateur Lin\'eaire  \\
   IN2P3-CNRS et Universit\'e de Paris-Sud }}
\end{center}
\vspace{\fill}
\begin{abstract}
\noindent
A review of the current status of the Cabibbo-Kobayashi-Maskawa matrix ($V_{CKM}$)
is presented and a special emphasis is put on the determination of
 the $\rho$ and $\eta$ parameters. From this study it follows that, in the Standard Model,
the $B^0_s$-$\overline{B^0_s}$ oscillation frequency, $\Delta m_s$, has to lie, with 68$\%$ C.L., 
between 6.5 ps$^{-1}$ and 15 ps$^{-1}$, and is below 21 ps$^{-1}$ at 95$\%$ C.L..
If the interest of measuring $\Delta m_s$ is underlined, 
%the necessity of an high luminosity run at the $\rm{Z^0}$ pole.
 the importance of a precise determination
of the B meson decay constant, $f_B$, is also stressed.
It is proposed to obtain a precise value for this parameter from an accurate measurement of 
$f_D$, the D meson decay constant, using results from lattice QCD
to relate D and B hadrons. A future Tau-Charm factory could accomplish this task. 
It is also shown that from the present measurements, assuming the validity of the Standard Model, an already accurate value of 
$sin 2\beta~=~ 0.67 ^{+0.12}_{-0.13}$ is obtained. Sofar no constraint can be obtained on $sin 2\alpha$.
The interest of having a direct measurement of $sin 2\alpha$ and $sin 2\beta$ at B-factories or at other facilities is reminded.
Finally constraints on SUSY parameters, in the framework of a given model,
obtained from a precise measurement of the CKM matrix elements are analyzed.
\end{abstract}

\vskip 2.0truecm
\vspace{\fill}
\begin{center}
To be submitted to  Physica Scripta
\end{center}
\begin{center}
 ( DELPHI 97-137 PHYS 724 and LAL-97-79)
\end{center}
 
\noindent
\vskip 3.0 truecm
\noindent
*~~~~~ Now at \'Ecole Polytechnique, LPNHE, Palaiseau  \\
\noindent
+~~~~~ on leave of absence of the INFN-Genova
\vspace{\fill}
\end{titlepage}

\setcounter{page}{1}    
\section{Introduction}
\label{sec:1}

The search for CP violation in B meson decays, at B-factories or high energy hadron colliders,
is usually justified by the quest to explain the asymmetry matter-antimatter in the Universe \cite{ref:sakha}.
The Standard Model CP violation, which is contained in the Cabibbo, Kobayashi, Maskawa matrix, fails by about 10 orders of
magnitude to account for the measured abundance of baryons \cite{ref:universe}. 
It is then necessary to invoke non-Standard Model
sources of CP violation and many possibilities exist. As an example, in the Supersymmetric extension of 
the Standard Model, additional phases are present but there is no guarantee that these phases
produce measurable effects in B decays or that they can really account for the cosmological excess of baryons.

In this paper, a pragmatic attitude has been adopted which consists in considering the limiting factors
on the precision of consistency tests of the Standard Model predictions and a special emphasis
has been placed on the control of uncertainties coming from the strong interaction operating in the 
non-perturbative regime.

In the Standard Model, weak interactions among quarks are codified in a 3$\times$3 unitary matrix: the V$_{CKM}$ matrix.
The existence of this matrix conveys the fact that quarks which participate to weak processes are a linear 
combination of mass eigenstates. 
%of those acting at the level of the strong interactions.
 The V$_{CKM}$ matrix can be parametrized in terms of four
parameters: $\lambda$, $A$, $\rho$ and $\eta$ and the Standard Model predicts relations between the different processes which
depend on these parameters; CP violation is one of those. The unitarity of the V$_{CKM}$ matrix can be visualized as a
triangle in the $\rho-\eta$ plane.
%Forgetting about cosmological implications, it remains that deviations from Standard Model expectations  
%can be detected from accurate measurements of processes which depend on the C.K.M. parameters.
%The Standard Model predict strict relations between these processes which may be violated if
%additional phenomena are present. 
%Since 1989, LEP has measured that there are only three families of quarks and leptons. This implies
%that the C.K.M. matrix is unitary and that the so-called unitarity triangle
%is really a triangle. The precise position of its summit, in the ($\rho, \eta$) plane is not
%predicted by the model.
Several quantities which depend on $\rho$ and $\eta$ can be measured and, if the Standard Model is 
correct, they must define compatible values for the two parameters, inside measurement
errors and theoretical uncertainties. Extensions of the Standard Model, as Supersymmetry, provide
different parametrizations, for the same observables, and additional parameters to $\rho$ and $\eta$.
The position of the summit of the unitarity triangle, given by the $\rho$ and $\eta$ coordinates,
will be different in this scenario. It has to be stressed that only the extentions of the Standard Model 
in which the flavour changes are controlled by the $V_{CKM}$ matrix are considered in this paper.
 
It is thus of interest to measure accurately as many observables as possible to have evidence
for physics outside the Standard Model. By the year 2000, HERA-B and B-factories are expected to measure
CP violation phases which are directly related to the angles $\alpha$ and $\beta$ of the unitarity triangle.
The aim of this paper is to review present constraints on $\rho$ and $\eta$ obtained from
existing measurements of $\mbox{B}^0_{d,s}-\overline{\mbox{B}^0_{d,s}}$ oscillations, CP violation
in the K system and $b \rightarrow u \ell \overline{\nu_{\ell}}$ transitions. With the same constraints,
and considering the expected improvements of current analyses and of lattice QCD evaluations, possible
uncertainties on $\rho$ and $\eta$ have been evaluated from future measurements.
 After the year 2000, the contribution from precise 
measurements at a Tau-Charm factory, has been also considered.
It is shown that a well defined region, in the ($\rho, \eta$) plane, can be obtained, independently of the
measurements of the angles $\alpha$ and $\beta$ of the unitarity triangle. 
As these angles are expected to be measured at the same time,
with a reasonable accuracy, it will be of interest to verify if the two approaches 
%select the same region
%in the $\rho-\eta$ plane.
give compatible results.

\section{ The Cabibbo-Kobayashi-Maskawa (CKM) matrix : an historical introduction }
\label{sec:2}

In the early 60's there were three quark flavours. The quark $u$ with charge +2/3, the quarks $d$ and $s$ 
of charge -1/3. By analogy with leptons it was suggested that the 
quarks were also organized into doublets and the existence of a new quark of charge 2/3 was
proposed \cite{ref:bjor}. An intense experimental activity on strange particles
shown, at the same time, that the absolute decay rate for $\Delta S$ =1 transitions 
was suppressed by a factor of about 20 as compared to $\Delta S$ =0 transitions. \\
In 1963 Cabibbo proposed \cite{ref:cabi} a model to account for this effect. In this model
the $d$ and $s$ quarks, involved in weak processes, are rotated by a mixing angle 
$\theta_{c}$ : the Cabibbo angle. The quarks are organized in a doublet : \\
\begin{equation}
\begin{array}{ccc}
\left( \begin{array}{c} u \\ d_c \end{array} \right)~=~
&
\left(  \begin{array}{c} u \\ d ~cos \theta_c + s ~sin \theta_c \end{array} \right)
\end{array}
\end{equation}
the small value of $\theta_c$ ($\simeq$ 0.22) is responsible for the suppression of
strange particle decays (the coupling being proportional to $sin^{2}\theta_{c}$). 
In this picture the slight suppression of $ n \rightarrow p e^{-} \overline{\nu_{e}}$ with respect to the rate of 
$ \mu^{-} \rightarrow e^{-} \nu_{\mu} \overline {\nu_{e}} $ is also explained by the fact 
that the coupling in the neutron decay is proportional to $cos^{2}\theta_{c}$.\\ In the other hand, in this model,
the neutral current coupling is of the form
\begin{equation}
u \overline{u} + d \overline{d} cos^2 \theta_c + s \overline{s} sin^2 \theta_c + (s \overline{d} +
d \overline{s} ) cos \theta_c sin \theta_c .
\end{equation}
The presence of the $(s \overline{d} + d \overline{s})$ term implies the existence of a flavour changing 
neutral current (FCNC). This was a serious problem for the Cabibbo model, since all the observed
neutral current processes were characterized by the selection rule $\Delta S$ = 0.\\
In 1970 Glashow, Iliopoulos and Maiani \cite{ref:gim} (GIM) proposed the introduction of a
new quark, named $c$, of charge 2/3 and the introduction of a new doublet of quarks formed by 
the $c$ quark and by a combination of the $s$ and $d$ quarks orthogonal to $d_c$:  \\
\begin{equation}
\begin{array}{ccc}
\left( \begin{array}{c} c \\ s_c \end{array} \right)~=~
&
\left( \begin{array}{c} c \\ s ~cos \theta_c - d ~sin \theta_c  \end{array} \right) .
\end{array}
\end{equation}
In this way the $(s \overline{d}+d \overline{s})$ term in the neutral current disappears.\\
The discovery of the charm quark in the form of $c \overline{c}$ bound states
\cite{ref:jpsi} and the observation of charmed particles decaying into 
strange particles \cite{ref:cleo} (the $c \overline{s}$ transitions which are proportional
to $cos^{2}\theta_c$ dominate over the $c \overline{d}$ transitions which are 
proportional to $sin^{2} \theta_c$) represent a tremendous triumph of
this picture.\\
The charge current can then be written: 
\begin{equation}
( \overline{u}  \overline{c} )  \gamma^{\mu} (1-\gamma_5) V 
\left( \begin{array}{c} d \\ s \end{array} \right)
\end{equation}
where $u,d,s,c$ are the mass eigenstates and V is defined as:
\begin{equation}
V =
\left(
\begin{array}{cc}
cos \theta_c   ~~~sin \theta_c  \\
-sin \theta_c  ~~~cos \theta_c 
\end{array}
\right).
\end{equation}
V is the Cabibbo unitary matrix which specifies the quark states which are involved in
weak interactions. This matrix express the fact that there is an arbitrary rotation, usually applied to the -1/3 charged quarks,
which is due to the mismatch between the strong and the weak eingenstates. \\
In 1975 the Mark I group at SPEAR discovered the third charged lepton : the $\tau$ \cite{ref:tau}. 
Two years later the fifth quark, the $b$, was also found at FNAL \cite{ref:fnal}. The indirect existence
for the top quark $t$ from the observation of $B_d^0-\overline{B_d^0}$ oscillations \cite{ref:bdbd} suggested the existence
of a heavier version of the doublets ($u$,$d$) and ($c$,$s$) (the $t$ quark has been recently discovered at Fermilab 
\cite{ref:top} in $p \overline{p}$ collisions) \\
The existence of three quark doublets was already proposed by Kobayashi and Maskawa in 1973 
\cite{ref:ckm} as a possible explanation of CP violation. Their proposal is a 
generalization of the Cabibbo rotation and implies that the weak flavour changing transitions
are described by a $3\times 3$ unitary matrix:
\begin{equation}
\begin{array}{ccccccccc}
\left ( \begin{array}{c} u \\ c \\ t \end{array} \right )
&
\rightarrow
&
V
&
\left ( \begin{array}{c} d \\ s \\ b \end{array} \right )
&
,
&
V =
&
\left ( \begin{array}{ccc} 
V_{ud} ~~ V_{us} ~~ V_{ub} \\
V_{cd} ~~ V_{cs} ~~ V_{cb} \\
V_{td} ~~ V_{ts} ~~ V_{tb}
\end{array} \right )
\end{array}.
\end{equation}
This matrix can be parametrized using three real parameters and
one phase which cannot be removed by redefining the quark field phases. This phase leads 
to the violation of the CP symmetry. In fact since CPT is a good symmetry for all quantum
field theories, the complexity of the Hamiltonian implies that the time reversal invariance
T and thus CP is violated. In this picture the Standard Model includes CP violation in a simple
way.\\
Several parametrizations of the $V_{CKM}$ matrix exist. \\
The standard parametrization is \cite{ref:pdg} :
\begin{equation}
\begin{array}{ccc}
V_{CKM} =
&
\left ( \begin{array}{cccc}
~~c_{12}c_{13}~~~~~~~~~~~~~~~~~~~~~~~~~  s_{12}c_{13} ~~~~~~~~~~~     s_{13}e^{-i\delta_{13}} \\
-s_{12}c_{23}-c_{12}s_{23}s_{13}e^{i\delta_{13}}  ~~~    c_{12}c_{23}-s_{12}s_{23}s_{13}e^{i\delta_{13}}~~~~    s_{23}c_{13} \\
~~s_{12}s_{23}-c_{12}c_{23}s_{13}e^{i\delta_{13}}   ~~~    -c_{12}s_{23}-s_{12}c_{23}s_{13}e^{i\delta_{13}} ~~~  c_{23}c_{13}
\end{array} \right )
\end{array}
\end{equation}
where $c_{ij}$ and $s_{ij}$ stand for $cos \theta_{ij}$ and  $sin \theta_{ij}$ respectively, the indices ${1,2,3}$ represent the 
generation label and $\delta_{13}$ is the phase. Experimentally there is a hierarchy in the magnitude of these elements. 
The parametrization proposed by Wolfenstein 
\cite{ref:wolf}, in terms of four parameters: $\lambda$, $A$, $\rho$ and $\eta$, is motivated by this observation. By definition
these parameters verify the following expressions:
$$ s_{12}= \lambda,~~s_{23}=A \lambda^2,~~s_{13}e^{-i\delta_{13}}=A \lambda^3 ( \rho-i \eta). $$
 The $V_{CKM}$ matrix elements can be expanded in powers of $\lambda$ and, without any prejudice on the possible
values for $A$, $\rho$ and $\eta$, expressions valid up to  $0(\lambda^6)$ have been obtained :
\begin{equation}
\begin{array}{cccc}
V_{CKM} =
&
\left ( \begin{array}{cccc}
1 - \frac{\lambda^{2}}{2} - \frac{\lambda^4}{8} ~~~~~~~~~~~~~~~~~~~~~~~~~~~~~~  \lambda ~~~~~~~~~~~~     
A \lambda^{3} (\rho - i \eta) \\
   - \lambda +\frac{A^2 \lambda^5}{2}(1-2 \rho) -i A^2 \lambda^5 \eta~~~~~~~~~~  1 - \frac{\lambda^{2}}{2}
   -\lambda^4(\frac{1}{8}+\frac{A^2}{2})
 ~~~~~~~~           A \lambda^{2}       \\
A \lambda^{3} (1 - (1-\frac{\lambda^2}{2})(\rho +i \eta)) ~~~ -A \lambda^{2}(1-\frac{\lambda^2}{2})(1 + \lambda^{2}(\rho +i \eta))
 ~~~    1-\frac{A^2 \lambda^4}{2}
\end{array} \right )
& 
%+ O(\lambda^{6}).
\end{array}
\label{eq:eq8}
\end{equation}
 $\eta$ accounts for the CP violating phase. This parametrization is still approximated but is completely adequated to the
 arguments developed in this paper. \\
This parametrization shows that the matrix is almost diagonal, namely that the coupling between
quarks of the same family is close to unity, and is decreasing as the separation 
between families increases. \\

\section{ The $A$ and $\lambda$ parameters }
\label{sec:3}

The $\lambda$ parameter has been measured using strange particle decays involving the $s \rightarrow u$ transitions.
It is obtained from the measurement of $K_{l3}$ decays : $K^{+} \rightarrow \pi^{0} \ell^{+} \nu_{e}$ or 
$K^{0}_{L} \rightarrow \pi^{-} \ell^{+} \nu_{e}$  and from hyperons semileptonic decays. The average of these
 measurements is \cite{ref:pdg}: \\
\begin{equation}
\mid V_{us} \mid = \lambda = 0.2205 \pm 0.0018 
\end{equation}
Studies of $c \rightarrow d$ transitions, in neutrino production of charmed particles and in charmed particle decays 
into non-strange states, give a compatible determination of $\lambda$ which is nearly ten times less precise \cite{ref:pdg}
\begin{equation}
\mid V_{us} \mid = \lambda = 0.224 \pm 0.016
\end{equation}
The average of the previous determinations gives the following value for $\lambda$:
\begin{equation}
\lambda = 0.2205 \pm 0.0018
\end{equation}

In the Wolfenstein parametrization the parameter $A$ is introduced and defined from the expression :
\begin{equation}
\mid V_{cb} \mid = A \lambda^{2}
\end{equation}
The value of $\mid V_{cb} \mid$  can be determined in two ways :
\begin{itemize}
\item
from LEP measurements of the inclusive lifetime, $\tau_b$, and of the semileptonic branching fraction
$(BR^b_{sl})$ of B hadrons. $\mid V_{cb} \mid$ can be obtained from the following expression :
\begin{equation}
\mid V_{cb} \mid^{2}  = \frac{BR_{sl}^{b}}{\gamma_{c} \tau_{b}}.
\label{eq:ocche}
\end{equation}
The term $\gamma_{c}$ includes the dependence on $m_b , m_c$ and contains also QCD corrections. 
An analogous definition is valid at the $\Upsilon(4S)$ using the measured branching ratio, $Br(\mbox{B}\rightarrow \ell^+ \nu X)$,
and the average $\mbox{B}^{+}/\mbox{B}_d^{0}$ lifetimes. 
The use of experimental data and of theoretical models \cite{ref:models} is subject to an intense discussion 
\cite{ref:bigi}. A detailed review can be found in \cite{ref:sl1} \cite{ref:sl2} \cite{ref:lalachi}.
From the values given in Table \ref{tab:tab1} and using the relation (\ref{eq:ocche}), $\mid V_{cb} \mid$ has been evaluated to be:
\begin{equation}
\mid V_{cb} \mid= 0.0405 \pm 0.0005 \pm 0.0030.
\end{equation}
The second quoted error reflects theoretical uncertainties on the parameter $\gamma_c$ and it has been overestimated according to
the authors of \cite{ref:varenna}.

\begin{table}
\begin{center}
\begin{tabular}{|c|c|c|}
\hline
                        Variables                   &             results             &      references         \\ \hline
$\tau_b$                                            &   $ 1.566 ~ \pm~  0.017 ~ ps$   & \cite{ref:life_wkg}     \\ \hline
$\tau_{B_{d}^{0}}$                                  &   $ 1.55  ~ \pm~  0.04  ~ ps$   & \cite{ref:life_wkg}     \\ \hline
$\tau_{B^+}$                                        &   $ 1.66  ~ \pm~  0.04  ~ ps$   & \cite{ref:life_wkg}     \\ \hline
$Br(b \rightarrow \ell^+ \nu X) (LEP)$              &   $ 11.22 ~ \pm ~ 0.21 \%   $   & \cite{ref:ewg}          \\ \hline
$Br(\mbox{B} \rightarrow \ell^{+} \nu X) (\Upsilon (4S))$  &   $ 10.43 ~ \pm ~ 0.24 \%   $   & \cite{ref:pdg}          \\ \hline
$\gamma_c$                                          &   $ 41    ~ \pm ~ 6 ~ps^{-1} $   & \cite{ref:bigi}         \\ \hline
\end{tabular}
\caption[]{ \it {Theoretical and experimental inputs used in the evaluation of $V_{cb}$. }}
\label{tab:tab1}
\end{center}
\end{table}

\item
from the interpretation of the measurement of the exclusive branching fraction, $BR(\mbox{B} \rightarrow \mbox{D}^{*} \ell \nu)$, 
in the HQET framework \cite{ref:hqet}. The measurements done at the $\Upsilon (4S)$ and at LEP \cite{ref:dstar} give:
\begin{equation}
 F(1) \mid V_{cb} \mid= 0.0357 \pm 0.0021 \pm 0.0014.
\end{equation}
Using for F(1), which deviates from unity because of strong interactions, the value \cite{ref:hqet}:
\begin{equation}
F(1) ~ = ~ 0.91 \pm 0.03 
\end{equation}
it follows:
\begin{equation}
\mid V_{cb} \mid= 0.0392 \pm 0.0027 \pm 0.0013.
\end{equation}
\end{itemize}
\noindent
Assuming that the theoretical uncertainties are uncorrelated and averaging the two previous determinations, 
$\mid V_{cb} \mid$ results to be :
\begin{equation}
\mid V_{cb} \mid= 0.0397 \pm 0.0020
\end{equation}
which corresponds to:
\begin{equation}
 A = 0.81 \pm 0.04
\end{equation}

\section  { The unitarity triangle}
\label{sec:4}

\begin{figure}
\begin{center}
{\epsfig{figure=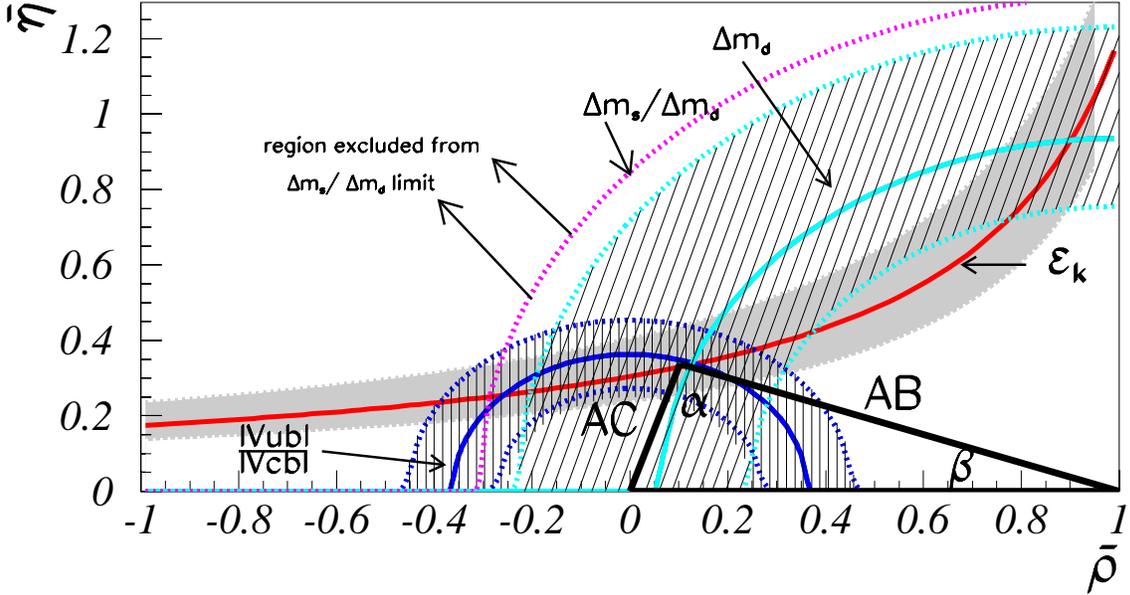,bbllx=30pt,bburx=489pt,bblly=6pt,bbury=251pt,height=8cm}}
\caption{ \it{The unitarity triangle in the ($\overline{\rho}$-$\overline{\eta}$) plane. The constraints on
$\Delta m_d$, $ \mid \epsilon_K \mid $ and $\frac{\left | V_{ub}\right |}{\left | V_{cb}\right |}$ are shown as bands 
corresponding to $\pm$ 1$\sigma$. The 95 $\%$ C.L. excluded region obtained from the limit on the ratio $\Delta m_s/\Delta m_d$ 
corresponds to the area situated on the left of the line.}}
\label{fig:tria}
\end{center}
\end{figure}

From the unitarity of the CKM matrix ( $ V V^{\dag} = 1 $) three independent equations relating its
elements can be written. In particular, in transitions involving $b$ quarks, the scalar 
product of the third column with the complex conjugate of the first row must vanish :
\begin{equation}
V_{ud}^{\ast} V_{ub}~+~ V_{cd}^{\ast} V_{cb}~+~ V_{td}^{\ast} V_{tb}~=~0
\label{eq:triangle}
\end{equation}
Using the parametrization given in eq. (\ref{eq:eq8}), and neglecting contributions of order $0(\lambda^7)$, 
the different terms, in this expression,
are respectively:
\begin{equation}
 V_{ud}^{\ast} V_{ub}~=~A \lambda^3(\overline{\rho}-i \overline{\eta}),~
 V_{cd}^{\ast} V_{cb}~=~-~A \lambda^3,~
 V_{td}^{\ast} V_{tb}~=~A \lambda^3(1-\overline{\rho}+i \overline{\eta})
\end{equation}
where the parameters $\overline{\rho}$ and $\overline{\eta}$ have been introduced \cite{ref:etabar}:
$$
\overline{\rho} = \rho ( 1-\frac{\lambda^2}{2} ) ~~~;~~~
\overline{\eta} = \eta ( 1-\frac{\lambda^2}{2} ). 
$$
The three expressions are proportional  to $A \lambda^3$ which can be factored out and 
the geometrical representation  of eq. (\ref{eq:triangle}), in the ($\overline{\rho}$-$\overline{\eta}$) plane,
is a triangle with summits at C(0,0), B(1,0) and A($\overline{\rho}$-$\overline{\eta}$) ( see Figure \ref{fig:tria} )
The following relations hold :
$$
\overline{AC} = \frac{1-\frac{\lambda^2}{2}}{\lambda}  \frac{\mid V_{ub} \mid}{\mid V_{cb} \mid} = 
{\sqrt{(\overline{\rho}^{2} + \overline{\eta}^{2})}} 
$$
$$
\overline{AB} = \frac{\mid V_{td} \mid}{\lambda \mid V_{cb} \mid} = \sqrt{((1-\overline{\rho})^{2} +\overline{ \eta}^{2})} 
$$
\begin{equation}
\overline{AB}  =  \frac{1-\frac{\lambda^2}{2}}{\lambda} \frac{\mid V_{td} \mid}{ \mid V_{ts} \mid} = 
\sqrt{((1-\overline{\rho})^{2} +\overline{ \eta}^{2})}
\end{equation}
these expressions are valid up to  $0(\lambda^4)$.\\
The Standard Model, with three families of quarks and leptons,
predicts that all measurements have to be consistent with the point A($\overline{\rho}-\overline{\eta}$).

\subsection{Charmless semileptonic decays of B mesons.}
\label{sec:41}

The presence of leptons above the kinematical limit for leptons produced in the decay
$\mbox{B} \rightarrow \mbox{D} \ell \overline{\nu_{\ell}}$, at the $\Upsilon$(4S),
is attributed to $b \rightarrow u \ell \overline{\nu_{\ell}}$ transitions.
As only a small fraction of the energy spectrum of these leptons is accessible experimentally,
there is, at present, a large systematic uncertainty in the modelling of these transitions to evaluate
the value of $\left | V_{ub} \right |$ \cite{ref:pdg}:
\begin{equation}
\frac{\left | V_{ub} \right |}{\left | V_{cb} \right |}~=~0.08 \pm 0.02.
\label{eq:vbubc} 
\end{equation}
This measurement corresponds to the following constraint in the ($\overline{\rho}$-$\overline{\eta}$) plane:

\begin{equation}
\sqrt{\overline{\rho}^{2} + \overline{\eta}^{2}}~=~0.35 \pm 0.09.
\end{equation}

\subsection{CP violation in the $\mbox{K}^0-\overline{\mbox{K}^0}$ system.}
\label{sec:42}

Indirect CP violation in the $\mbox{K}^0-\overline{\mbox{K}^0}$ system is usually expressed
in terms of the $\mid \epsilon_K \mid$ parameter which is the fraction of the CP violating component in the 
mass eigenstates.
%Considering that at $\lambda^{4}$ order the $\mid V_{cs} \mid$ element is 
%$ ( 1 - \lambda^{2}/2 - i A \lambda^{4} \eta )$ , $ \mid \epsilon \mid $ can be expressed by:
Replacing $\left | V_{cs} \right |$ by its expression in terms of the Wolfenstein parameters,
the following equation is obtained, having neglected $0(\lambda^4)$ terms:  
\begin{eqnarray}
\mid \epsilon_K \mid~ =~ C_{\epsilon}~ \mbox{B}_K~A^2 \lambda^6 \overline{\eta} [-\eta _1 
(1-\frac{\lambda^2}{2})S(x_c)+A^2 \lambda^4(1-\overline{\rho}-
(\overline{\rho}^2+\overline{\eta}^2)\lambda^2) \eta _2 S(x_t) \nonumber \\
+ \eta _3 S(x_c,x_t)].
\label{eq:epsilon}
\end{eqnarray} 

Another contribution, proportional to the ratio $ \xi = Im(A(K \rightarrow \pi \pi)_{I=0}) / Re(A(K \rightarrow \pi \pi)_{I=0})$,
which is at most a 2$\%$ correction to $\epsilon_K$ \cite{ref:buras0} has been also neglected.
The constant $C_{\epsilon}~=~\frac{G_F^2 f_K^2 m_K m_W^2}{6\sqrt{2} \pi^2 \Delta m_K}$
is equal to $3.85 \times 10^4$ (see Table \ref{tab:a}).
The $\eta_i$ parameters include perturbative QCD corrections which have been evaluated
recently at the next to leading order \cite{ref:buras1} \cite{ref:buras2}:
\begin{equation}
\eta_1~=~1.38 \pm 0.53,~~~\eta_2~=~0.574 \pm 0.004,~~~\eta_3~=~0.47 \pm 0.04.
\end{equation}
The expressions for $S(x_c)$ and $S(x_c,x_t)$ depend on $x_{q}=\frac{m_q^2}{m_W^2}$.
An important theoretical uncertainty in eq. (\ref{eq:epsilon}) comes from the ``bag'' parameter $B_K$ which is of non-perturbative
QCD origin. 
%Analyses using lattice QCD methods \cite{ref:bk1} are in relatively good
%agreement with those using the 1/N approach \cite{ref:bk2}.
From recent lattice QCD evaluations the following value can be quoted \cite{ref:gerard}: 
\begin{equation}
B_K(2 GeV)~=~0.66 \pm 0.02 \pm 0.06.
\end{equation}
The first error corresponds to the present uncertainty of this evaluation inside
the quenched approximation and the second error is related to the quenched approximation
itself. The scale-invariant value for $B_K$ is then evaluated, so that it will be 
consistent with the conventions used to obtain the $\eta_i$ parameters \cite{ref:buras2}:
\begin{equation}
B_K~=~(\alpha^{(3)}_s(\mu))^{-\frac{2}{9}}(1 + \frac{\alpha^{(3)}_s(\mu)}{4 \pi} J_3)~B_K(\mu).
\end{equation}
In this expression, $J_3~=~\frac{307}{162}$, $\Lambda_{QCD}^{(3)}~=~371~MeV$
and the following value is obtained for $B_K$ using $\mu$ = 2 GeV:
\begin{equation}
B_K~=~0.90 \pm 0.09.
\end{equation}

\subsection{$\mbox{B}^0_d - \overline{\mbox{B}^0_d}$ oscillations.}
\label{sec:43}

In the Standard model, the mass difference between the mass eigenstates in the 
$\mbox{B}^0_d - \overline{\mbox{B}^0_d}$ system can be expressed as:
\begin{equation}
%\Delta m_d~=~\frac{G_F^2}{6 \pi^2} \mid V_{tb} \mid^2 \mid V_{td} \mid^2
%m_t^2 m_{B_d} f^2_{B_d} B_{B_d} \eta_B F(x_t).
\Delta m_d~=~\frac{g^4}{192 m_{W}^{2} \pi^2} \mid V_{tb} \mid^2 \mid V_{td} \mid^2
m_{B_d} f^2_{B_d} B_{B_d} \eta_B x_t F(x_t),
\label{eq:deltamd1}
\end{equation}

where $g^2=\frac{G_F}{\sqrt{2}} 8 m_W^2$. In this expression only terms in which the top quark contributes have been retained
because they dominate. The mixed term with top and charm quark is almost 600 times smaller than the dominant one; the one
involving only charm quarks is suppressed by an extra factor ten \cite{ref:pascal}.
The function $F(x_t)$ is given by: 
\begin{equation}
F(x_t)=\frac{1}{4}+\frac{9}{4}\frac{1}{(1-x_t)}-\frac{3}{2} \frac{1}{(1-x_t)^2}
-\frac{3}{2} \frac{x_t^2}{(1-x_t)^3}\ln x_t
\label{eq:Ft}
\end{equation}
where $x_t$ = $(m^2_{t}/m^2_{W})$. $F(x_t)$ has a smooth variation with $x_t$ and is equal to 0.54, for $m_t = 180~ GeV/c^2$. 
The scale for the evaluation of perturbative QCD corrections
entering into $\eta_B$ and the running of the $t$ quark mass have to be defined in
a consistent way \cite{ref:buras2}. The measured value of the pole top quark
mass obtained by CDF and D0 collaborations ($m_t^{pole}~=~175 \pm 6~ GeV/c^2$ \cite{ref:top}) has to be
corrected downwards by $(7 \pm 1) GeV/c^{2}$. The following values have been used:
\begin{equation}
m_t~=~168 \pm 6~ GeV/c^2,~~~\eta_B~=~0.55 \pm 0.01.
\end{equation}
The dominant uncertainties in eq. (\ref{eq:deltamd1}) comes from 
the evaluation of the B meson decay constant $f_{B_d}$ and of the ``bag'' parameter $B_{B_d}$.
There is a vast literature on lattice QCD calculations of $f_{B}$. From the reviews \cite{ref:fbrev} based on
recent studies \cite{ref:fbstu} , the preferred value is :
\begin{equation}
f_{B_d} ~=~175 \pm 25 \pm 30~MeV
\label{eq:fb}
\end{equation}
where the first uncertainty is obtained using the quenched approximation and the second uncertainty accounts
for the use of this approximation itself. 
 These results are compatible with those
obtained using sum rules technique \cite{ref:sumrules}, the relativistic quark model \cite{ref:relativistic} and the
QCD potential model \cite{ref:qcd_potential}.\\
The bag factor $B_{B_d}$ has been evaluated using lattice QCD in a way which is consistent with the result quoted for $B_K$.
%It is consistent with previous results obtained in lattice QCD \cite{ref:bbag1} or using sum rules \cite{ref:bbag2} techniques.
At the 2 GeV scale, the following value from lattice QCD is obtained: 
\begin{equation}
B_{B_d}(2~GeV) = 0.95 \pm 0.07 \pm 0.09,
\label{eq:bb}
\end{equation}
where the errors have the same meaning as in eq. (\ref{eq:fb}).
The scale-invariant value for $B_{B_d}$ has been obtained using the following expression \cite{ref:buras1}:
\begin{equation}
B_{B_d}~=~(\alpha^{(5)}_s(\mu))^{-\frac{6}{23}}(1 + \frac{\alpha^{(5)}_s(\mu)}{4 \pi} J_5)~B_{B_d}(\mu),
\end{equation}
in which $J_5=\frac{5165}{3174}$ and gives:\\
\begin{equation}
 B_{B_d}~=~1.36 \pm 0.16.
\label{eq:bb11}
\end{equation}
Due to the dependence in $f_{B_d}^2 \times B_{B_d}$ in the expression of $\Delta m_d$ (eq. \ref{eq:deltamd1} ),
 the uncertainty on $f_{B_d}$ dominates over the one on $B_{B_d}$. 
The variable $f_{B_d} \sqrt {B_{B_d}}$ will be used often in the following and from expressions (\ref{eq:fb}) and (\ref{eq:bb11})
its value is equal to:
\begin{equation} 
f_{B_d} \sqrt{B_{B_d}}~=~200 \pm 50~MeV
\label{eq:fbfb}
\end{equation} 

The relative uncertainty on the square of this variable, which enters into expression (\ref{eq:deltamd1}), is then at the level of
$50 \%$. Using the parametrization given in eq. (\ref{eq:deltamd1}) the expression for $\Delta m_d$ becomes:
\begin{equation}
\Delta m_d~=\frac{g^4}{192 m_{W}^{2} \pi^2}~A^2 \lambda^6[(1-\overline{\rho})^2+\overline{\eta}^2]
m_{B_d} f^2_{B_d} B_{B_d} \eta_B x_t F(x_t).
\label{eq:deltamd2}
\end{equation}
which gives a constraint on the AB side of the triangle, corresponding to a circle centered on B, in the 
($\overline{\rho}$-$\overline{\eta}$) plane.

\subsection{$\mbox{B}^0_s - \overline{\mbox{B}^0_s}$ oscillations.}
\label{sec:44}

The ratio between the Standard Model expectations for $\Delta m_d$ and $\Delta m_s$ is
given by the following expression :
\begin{equation}
\frac{\Delta m_d}{\Delta m_s}~=~\frac{ m_{B_d} f^2_{B_d} B_{B_d} \eta_{B_d}}
{ m_{B_s} f^2_{B_s} B_{B_s} \eta_{B_s}} \frac{\left | V_{td} \right |^2}{\left | V_{ts} \right |^2}
\label{eq:ratiodms}
\end{equation}
Neglecting terms of order $0(\lambda^4)$, $\mid V_{ts} \mid$ is independent of $\rho$ and $\eta$ 
and is equal to $\mid V_{cb} \mid \times ({1-\frac{\lambda^2}{2}})$. 
A measurement of the ratio $\frac{\Delta m_d}{\Delta m_s} $
gives the same type of constraint, in the $\overline{\rho}$-$\overline{\eta}$ plane,
 as a measurement of $\Delta m_d$ because it is also proportional to
the length of the side AB of the unitarity triangle. This ratio is expected to be less dependent on the absolute 
values  of $f_B$ and $B_B$, and lattice QCD predicts \cite{ref:bbag1}:
\begin{equation}
 \xi = \frac{ f_{B_s} \sqrt{B_{B_s}}}{ f_{B_d} \sqrt{B_{B_d}}}~=~1.17 \pm 0.06 \pm 0.12.
\label{eq:rdbds}
\end{equation}
The first error corresponds to the present uncertainty of this evaluation inside
the quenched approximation and the second error is related to the quenched approximation itself.
 
The present limit on $\Delta m_s$ allows to restrict the accessible domain for the
$\overline{\rho}$ and $\overline{\eta}$ parameters. \\
Measurements of $\Delta m_d$, $\left | V_{ub} \right |$ and $\mid \epsilon_K \mid$ can be used also to define the most probable 
region for $\Delta m_s$ and ${ f_{B_d} \sqrt{B_{B_d}}}$.

\section{Present constraints on the $\overline{\rho}$ and $\overline{\eta}$ parameters.}
\label{sec:5}

\subsection{ The methods }
\label{sec:51}

The constraints on  $\overline{\rho}$ and $\overline{\eta}$  have been obtained in two ways.\\
These parameters can be determined by fitting their values from the three expressions given in 
$(\ref{eq:vbubc})$, $(\ref{eq:epsilon})$, $(\ref{eq:deltamd2})$, $(\ref{eq:ratiodms})$ for, respectively,
$  \frac{\left | V_{ub} \right |}{\left | V_{cb} \right |}$, $ \mid \epsilon_K \mid$, $\Delta m_d$, $\frac{\Delta m_d}{\Delta m_s}$
and the corresponding measurements. Extra constraints have been added on $A$, $m_{t}$, $B_K$, 
$f_{B_d}\sqrt{B_{B_d}}$, $\xi$ and the QCD parameters entering in the expression of $ \mid \epsilon_K \mid$, whereas uncertainties on 
the other parameters have been neglected. The list of parameters used in this analysis is given in Table \ref{tab:a}. \\
An equivalent approach has been developed which consists in building up the two dimensional probability distribution 
for $\overline{\rho}$ and $\overline{\eta}$. This is done in the following way.
\begin{table}
\begin{center}
\begin{tabular}{|c|c|c|c|}
\hline
 today's values & references & ``year 2000'' & status \\
\hline
 A  ~=~0.81 $\pm$ 0.04 & this paper & $\pm 0.025$ & varied \\
 $\frac{\left | V_{ub} \right |}{\left | V_{cb} \right |} ~=~0.08 \pm 0.02 $ & \cite{ref:pdg} & $\pm 0.01$ & varied \\
 $\overline{m_c}(m_c)~=~(1.3 \pm 0.1)~GeV/c^2 $  & \cite{ref:pdg} &  & varied \\
 $f_K~=~0.161~GeV/c^2$  & \cite{ref:pdg} &  & fixed \\
 $\Delta m_K~=~(0.5333 \pm 0.0027) \times 10^{-2} ~ps^{-1}$  & \cite{ref:pdg} & & fixed \\
 $\Delta m_d~=~(0.469 \pm 0.019) ~ps^{-1}$  & \cite{ref:osciwg} & $\pm 0.015ps^{-1}$ & varied \\
 $\Delta m_s > 8.0 ps^{-1} ~{\rm at}~95 \% C.L.$  & \cite{ref:osciwg} & $12.5ps^{-1} ~{\rm at}~95 \% C.L$ & varied \\
 $\mid \epsilon_K \mid~=~( 2.258\pm 0.0018) \times 10^{-3}$  & \cite{ref:pdg} & & fixed \\
 $ \overline{m_t}(m_t)~=~(168 \pm 6) ~GeV/c^2$  & \cite{ref:top} & $\pm 5~GeV/c^2 $ & varied \\
 $B_K~=~0.90 \pm 0.09$  & this paper & $\pm 0.05$ & varied \\
 $ f_{B_d} \sqrt{B_{B_d}}~=~(200 \pm 50)~MeV$  & this paper & $  \pm 30 ~MeV$  & varied \\
%                                               &            & ($\pm 5 ~MeV$ in SCENARIO II) & varied \\
 $\frac{ f_{B_s} \sqrt{B_{B_s}}}{ f_{B_d} \sqrt{B_{B_d}}}~=~1.17 \pm 0.13$ & this paper & $\pm 0.06$ & varied \\
 $\eta_1~=~1.38 \pm 0.53$ & \cite{ref:buras1},\cite{ref:buras2} & & varied \\
 $\eta_2~=~0.574 \pm 0.004$ & \cite{ref:buras1},\cite{ref:buras2} & & fixed \\
 $\eta_3~=~0.47 \pm 0.04$ & \cite{ref:buras1},\cite{ref:buras2} & & varied \\
 $\eta_B~=~0.55 \pm 0.01$ & \cite{ref:buras1} & & fixed \\
 $ G_F~=~(1.16639 \pm 0.00002)\times 10^{-5} GeV^{-2}$  & \cite{ref:pdg} & & fixed \\
 $ m_{W}~ = ~80.33 \pm 0.15 ~GeV/c^2 $  & \cite{ref:pdg} & & fixed \\ 
 $ m_{B^0_d}~ = ~5.279 \pm 0.002 ~GeV/c^2 $ & \cite{ref:pdg} & & fixed \\ 
 $ m_{B^0_s}~ = ~5.375 \pm 0.006 ~GeV/c^2 $ & \cite{ref:pdg} & & fixed \\ 
 $ m_K~  = ~ 0.49767 \pm 0.00003 ~GeV/c^2 $  & \cite{ref:pdg} & & fixed \\ 
 $ \lambda~ = ~0.2205 \pm 0.0018$ &  \cite{ref:pdg} & & fixed \\ 
\hline
\end{tabular}
\caption[]{ \it {Values of the parameters entering into the expressions for $\Delta m_d$ (eq. \ref{eq:deltamd2}), 
$\left | \epsilon_K \right |$ (eq. \ref{eq:epsilon}), $\frac{\left | V_{ub}\right |}{\left | V_{cb} \right |}$ (eq. \ref{eq:vbubc})
 and $\Delta m_d/\Delta m_s$ (eq. \ref{eq:ratiodms}). The column ,"year 2000", gives the expected uncertainties 
on the parameters for which an improvement is forseen before that time.}}
\label{tab:a}
\end{center}
\end{table}

\begin{itemize}
\item
a point uniformly distributed in the ($\overline{\rho}-\overline{\eta}$) plane, is chosen,
\item
values for the different parameters entering into the equations of constraints
are obtained using random generations extracted from Gaussian distributions
\item
 the predicted values for the four quantities,
$ \frac{\left | V_{ub} \right |}{\left | V_{cb} \right |}$, $ \mid \epsilon_K \mid $, $\Delta m_d$ and 
$\frac{\Delta m_d}{\Delta m_s}$
are then obtained and compared with present measurements. A weight is computed
assuming that the measurements have Gaussian errors. As an example,
the weight corresponding to the constraint provided by the
measurement of $\Delta m_d$ is:
\begin{equation}
w^i(\Delta m_d) \propto \exp{-((\Delta m_d(meas.) - \Delta m_d(A^i,\rho^i,\eta^i,f^i_{B_d},
B^i_{B_d},m^i_t,\eta_{B_d}^i))^2/2\sigma(\Delta m_d)^2})
\end{equation}
where $\sigma(\Delta m_d)$ is the measurement error on $\Delta m_d$.
The final weight is equal to the product of the four weights.
% obtained from
%the three constraints.
\item
the sum of all weights, over the ($\overline{\rho}-\overline{\eta}$) plane, is normalized to unity and contours 
corresponding to 68$\%$ and 95$\%$ confidence levels have been defined.
\end{itemize}

It has been verified that the two approaches give equivalent results by comparing these contours
with those given by the fitting procedure. The second approach allows to construct the probability
distribution for several parameters of interest such as $\Delta m_s$, $sin 2\alpha$ and $sin 2\beta$.
In this approach non-Gaussian distributions for the errors on the experimental measurements of the constraints and on the
parameters entering into the equations of the constraints can be used. An example is given in (Section ~\ref{sec:522}). 

\subsection{Results with present measurements. }
\label{sec:52}

\subsubsection{Measured values of the $\overline{\rho}$ and $\overline{\eta}$ parameters:}
\label{sec:521}

\begin{figure}
\begin{center}
{\epsfig{figure=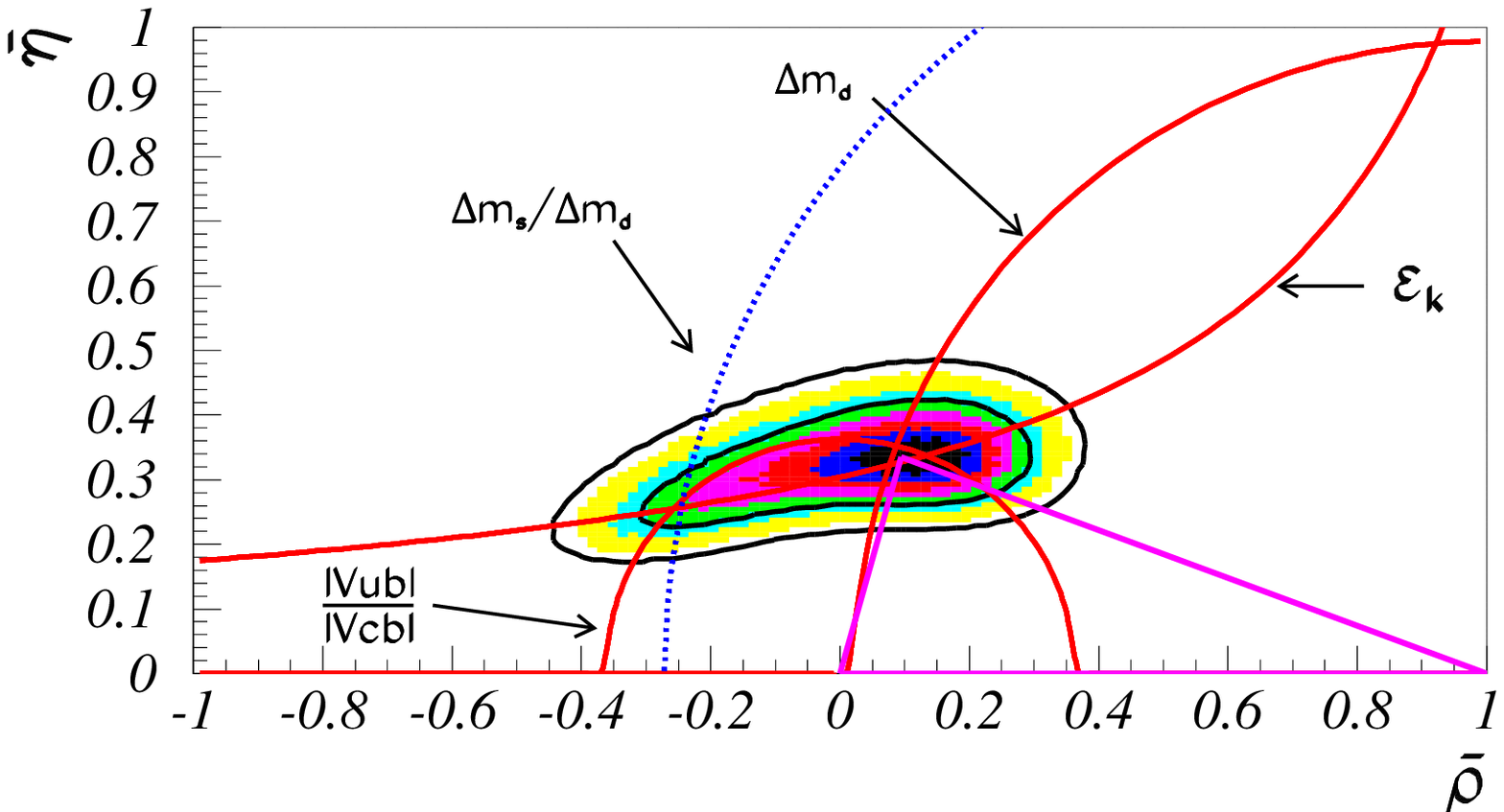,bbllx=30pt,bburx=489pt,bblly=6pt,bbury=251pt,height=8cm}}
\caption{ \it{The allowed region for $\overline{\rho}$ and $\overline{\eta}$ using the parameters listed in Table \ref{tab:a}.
The contours at 68$\%$ and 95 $\%$ are shown. The continous lines correspond to the constraints obtained from
the measurements of $  \frac{\left | V_{ub} \right |}{\left | V_{cb} \right |} $, $  \mid \epsilon_K  \mid $ and $\Delta m_d$.
The dotted curve corresponds to the 95 $\%$ C.L. upper limit obtained from the experimental limit on
$\Delta m_s$. This last constraint has not been included when
determining the allowed region. The position of the apex of the triangle corresponds to the most probable value which
co\"\i ncides with the fitted value given in (\ref{eq:result1}). }}
\label{fig:hari_1}
\end{center}
\end{figure}

The contours corresponding to 68$\%$ and 95$\%$ confidence levels are shown in Figure ~\ref{fig:hari_1}.
The result of the fit gives :
\begin{equation}
\overline{\rho} = 0.10 ^{+0.13}_{-0.38} ~~ ; ~~\overline{\eta} = 0.33 ^{+0.06}_{-0.09}  
\label{eq:result1}
\end{equation}
It has been verified that, when using the same set of values for the parameters listed in Table \ref{tab:a} as those
given in \cite{ref:alibaba}, the fitted values for $\overline{\rho}$ and $\overline{\eta}$ and their corresponding
errors were also similar.

The different sources of error which determine the size of the allowed region are examined in the following. 
Simplified expressions for the uncertainties on $\overline{\rho}$ and $\overline{\eta}$
have been obtained assuming that these two parameters are determined by the measurements of 
$\Delta m_d$ and $\mid \epsilon_K \mid$ respectively. The expressions are:
\begin{equation}
\sigma_{\overline{\rho}} \simeq \frac{\sigma_A}{A} \oplus 0.75 \frac{\sigma_{m_t}}{m_t}
\oplus 0.5 \frac{\sigma_{\Delta m_d}}{\Delta m_d} \oplus  \frac{\sigma_{f_{B_d} \sqrt{B_{B_d}}}}{f_{B_d} \sqrt{B_{B_d}}}
\oplus \sigma_{QCD}
\end{equation}

\begin{equation}
\frac{\sigma_{\overline{\eta}}}{\overline{\eta}} \simeq 3.3 \frac{\sigma_A}{A} \oplus \frac{\sigma_{m_t}}{m_t}
\oplus 0.7 \frac{\sigma_{m_c}}{m_c} \oplus  \frac{\sigma_{B_K}}{B_K}
\oplus \sigma_{QCD}
\end{equation}
 
The last mentioned contribution in these expressions comes from the uncertainty in the evaluation
of effects from strong interactions at the next to leading order.
Using the central values for the parameters listed in Table \ref{tab:a} and their corresponding uncertainties,
the expected error on $\overline{\rho}$ becomes:
$$
\sigma_{\overline{\rho}} \simeq 0.050 \oplus 0.027 \oplus 0.020 \oplus 0.25 \oplus 0.01
$$
The error on $\overline{\rho}$ is almost entirely due to the uncertainty on $f_{B_d} \sqrt{B_{B_d}}$ which is of theoretical 
origin (\ref{eq:fbfb}). Its present value does not allow to profit from the accurate measurement of the
$\Delta m_d$ parameter, which is known at the level of 4\% relative error ( see Table \ref{tab:a} ). The possibility of 
having a more precise determination of $f_{B_d}$ is considered in the section dedicated to the perspectives (Section ~\ref{sec:6}). 
The second important contribution to the error on $\overline{\rho}$ is coming from the uncertainty on the $A$ parameter extracted 
from the measurement of $\left | V_{cb} \right |$.

For the $\overline{\eta}$ parameter, the corresponding result is obtained:
$$
\frac{\sigma_{\overline{\eta}}}{\overline{\eta}} \simeq 0.16 \oplus 0.036 \oplus 0.054 \oplus 0.10 \oplus 0.055
$$

The uncertainty on $\overline{\eta}$ is governed by the theoretical uncertainty on $B_K$ and by the measurement error on
$ A$. The error on $A$ plays an important role since, for $\rho \simeq 0$, eq. (\ref{eq:epsilon})
tells  that: $ \mid \epsilon_K \mid \simeq B_K A^{4}$. 

In the considerations, developped in the second part of this paragraph, the effect 
from the measurement of $ \frac{\mid V_{ub} \mid }{\mid V_{cb} \mid} $, which can further reduce the accessible region, has been
neglected.

\subsubsection {Measured values for $\overline{\rho}$ and $\overline{\eta}$ when including the 
current limit on $\Delta m_s$.}
\label{sec:522}

\begin{figure}
\begin{center}
{\epsfig{figure=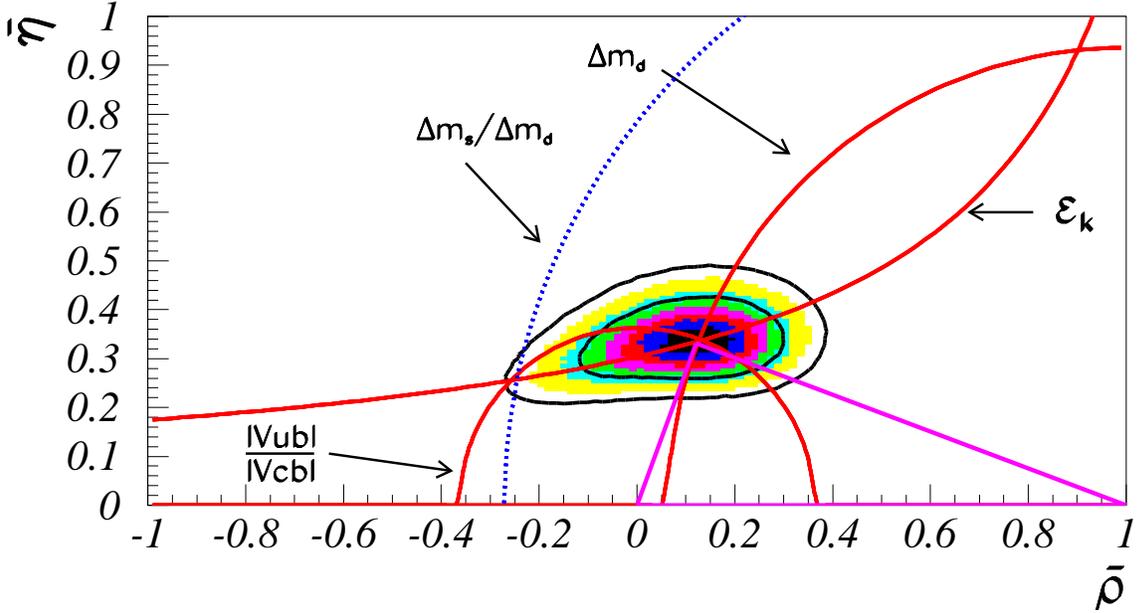,bbllx=30pt,bburx=489pt,bblly=6pt,bbury=251pt,height=8cm}}
\caption{ \it{ The allowed region for $\overline{\rho}$ and $\overline{\eta}$ using the parameters listed in Table \ref{tab:a}. 
The contours at 
68$\%$ and 95 $\%$ are shown. The full lines correspond to the central values of the constraints given by
the measurements of  $  \frac{\left | V_{ub} \right |}{\left | V_{cb} \right |} $, $  \mid \epsilon_K  \mid $ and $\Delta m_d$. 
The dotted curve corresponds to the 95 $\%$ C.L. upper limit obtained from the experimental limit on
$\Delta m_s$. This last constraint has been included following the procedure discussed
in Section \ref{sec:522}. The ratio $\xi$ has been also varied according to eq. (\ref{eq:rdbds}). }}
\label{fig:anno1997_dms}
\end{center}
\end{figure}

As mentioned in section ~\ref{sec:44}  a possibility to reduce the uncertainty on $\overline{\rho}$ is the measurement of the
 ratio $\frac{\left | \Delta m_d \right |}{\left | \Delta m_s \right |}$.
 The search for $\mbox{B}^0_s - \overline{\mbox{B}^0_s}$ oscillations has been the object of an intense 
activity in the last two years. No measurement is available sofar. The best limit comes from the 
combination of ALEPH, DELPHI and OPAL results \cite{ref:osciwg} :

\begin{equation}
\Delta m_s > 8.0 ps^{-1} ~~~~~~{\rm at}~~95 \% C.L.
\end{equation}
This limit has been obtained in the framework of the amplitude method \cite{ref:amplitude} which 
consists in measuring, for each value of the frequency $\Delta m_s$, an amplitude $a$ and its error $\sigma(a)$.
%($A=1$ means $\Delta m_s$ compatible with the true oscillation frequency $\Delta m_s$). A $95 \%$ C.L. is defined by the formula:
The parameter $a$ is introduced in the time evolution of pure $\mbox{B}^0_s$ or $\overline{\mbox{B}^0_s}$ states so that the
value $a=1$ corresponds to a genuine signal for oscillation:
$$
{\cal P}(\mbox{B}^0_s\rightarrow (\mbox{B}^0_s,~\overline{\mbox{B}^0_s}))~
=~\frac{1}{2 \tau_{s}} e^{- \frac{t}{\tau_{s}}} \times
 ( 1 \pm a~cos ({\Delta m_s t} ) )
$$
A randomly distributed decay time distribution which is expected for very large values of $\Delta m_s$, corresponds to $a=0$ and 
the 95$\%$ C.L. excluded region for $\Delta m_s$ is obtained in evaluating the probability that, in at most 5$\%$ of the cases, 
the observed amplitude is compatible with the value $a=1$, this corresponds to the condition: 
$$
a(\Delta m_s) + 1.645\sigma(a(\Delta m_s))~<~1.
$$
This limit on $\Delta m_s$ allows to cut the left side part of the $\overline{\rho}-\overline{\eta}$ 
region (Figure~\ref{fig:anno1997_dms}). 
However the set of measurements $a(\Delta m_s)$ contains more information than the $95 \%$ C.L. limit. It is possible
to build a $\chi^2$, which quantifies the compatibility of $a(\Delta m_s)$ with the value $a=1$, defined as:
$$  
\chi^2(\Delta m_s)  = {(a(\Delta m_s)-1)^2 \over \sigma^2(a(\Delta m_s))^2}
$$ 
and which can be used as a constraint.\\
The new allowed region for the $\overline{\rho}$ and $\overline{\eta}$ parameters is shown in Figure \ref{fig:anno1997_dms}.

The result of the fit gives :

\begin{equation}
\overline{\rho} = 0.11 ^{+0.13}_{-0.25} ~~ ; ~~ \overline{\eta} = 0.33 ^{+0.06}_{-0.06}
\label{eq:res1}
\end{equation}

\begin{figure}
\begin{center}
{\epsfig{figure=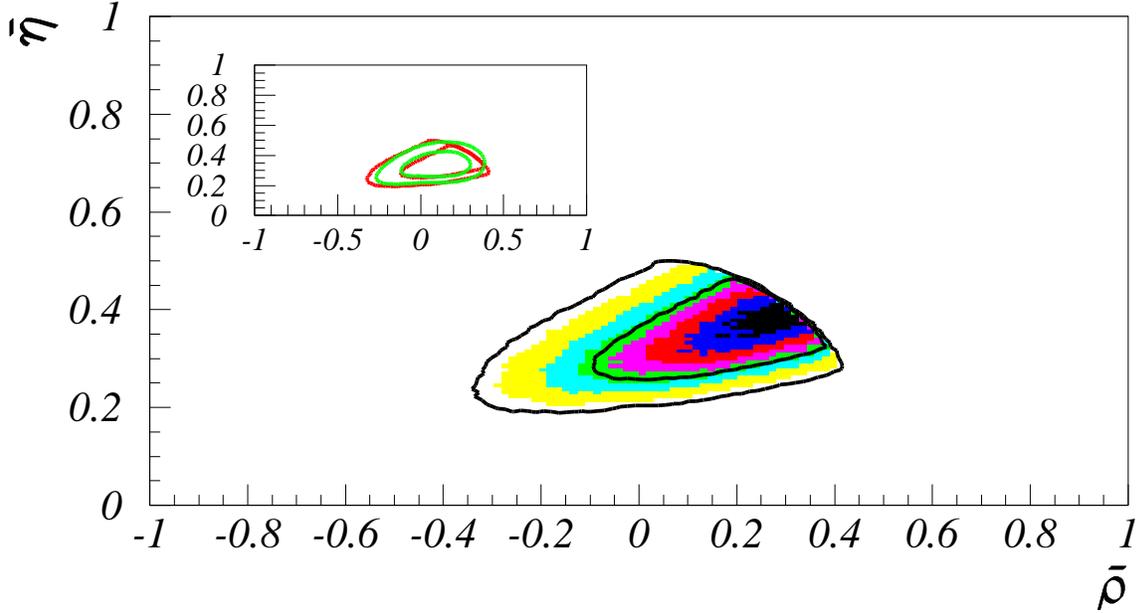,bbllx=30pt,bburx=489pt,bblly=6pt,bbury=251pt,height=8cm}}
\caption{ \it{ The allowed region for $\overline{\rho}$ and $\overline{\eta}$ using the parameters listed in Table \ref{tab:a}. 
Flat distributions have been used for $\frac{\left | V_{ub}\right |}{\left | V_{cb}\right |}$, $B_K$, 
$f_B \sqrt{B_{B}}$ and $\xi$ which have the same variance as the previously Gaussian distributions. The contours at 
68$\%$ and 95 $\%$ are shown. The inset shows the comparison between the contours at 68$\%$ and 95 $\%$ for Gaussian
distributions ( clearer contours ) and for flat distributions ( darker countours ).}}
\label{fig:anno1997_dms_piatto}
\end{center}
\end{figure}

It can be noticed that the present limit on $\Delta m_s$ gives a reduction by a factor of 1.5 
on the negative error of $\overline{\rho}$.\\
A similar fit has been performed using flat distributions for the quantities in which the systematic error is dominant. 
These quantities are :$\frac{\left | V_{ub}\right |}{\left | V_{cb}\right |}$, $B_K$, $f_B \sqrt{B_{B}}$ and $\xi$.
The considered variation for these distributions have been chosen such that the new, flat, and the previous, Gaussian,
distributions have the same variance. The new allowed region for the $\overline{\rho}$ and $\overline{\eta}$ parameters is shown 
in Figure \ref{fig:anno1997_dms_piatto}.

The comparison between the contours at 68$\%$ and 95 $\%$ of this new allowed region with those obtained with only Gaussian
distributions ( Figure \ref{fig:anno1997_dms} ) is shown in the inset of Figure \ref{fig:anno1997_dms_piatto}. The difference is
small and is essentially due to the uncertainty on $\frac{\left | V_{ub}\right |}{\left | V_{cb}\right |}$.

\subsubsection{ Simultaneous measurement of $\overline{\rho}$-$\overline{\eta}$ and $f_{B_{d}}\sqrt{{B}_{B_{d}}}$ }
\label{sec:523}

Instead of using  $f_{B_d}\sqrt{B_{B_d}}$ as an external constraint, this quantity can be also fitted together with 
the $\overline{\rho}$ and the $\overline{\eta}$ parameters. 
The result is:
\begin{eqnarray}
\overline{\rho} = 0.12 ^{+0.15}_{-0.26} ~~ ; ~~ \overline{\eta} = 0.33 ^{+0.06}_{-0.07}   \nonumber \\
 ~~~~~~~~~~~~~~~~~~~~~f_{B_d}\sqrt{B_{B_d}} = 208 ^{+30}_{-40} MeV
\label{eq:fb_mes}
\end{eqnarray}
If, in addition, the external constraint on $f_{B_d}\sqrt{B_{B_d}}$ is imposed, the result on the  $\overline{\rho}$ and 
$\overline{\eta}$ parameters is the one given in (\ref{eq:res1}) and $f_{B_d}\sqrt{B_{B_d}} = 205 ^{+26}_{-35} MeV$.
This result implies that the constraint coming from  the measurement of $\Delta m_d$ 
will bring information only if the uncertainty on $f_{B_d}\sqrt{B_{B_d}}$ can be reduced to 30 MeV or below.

\subsubsection{ $\Delta m_s$ probability distribution.}
\label{sec:524}

\begin{figure}
\begin{center}
{\epsfig{figure=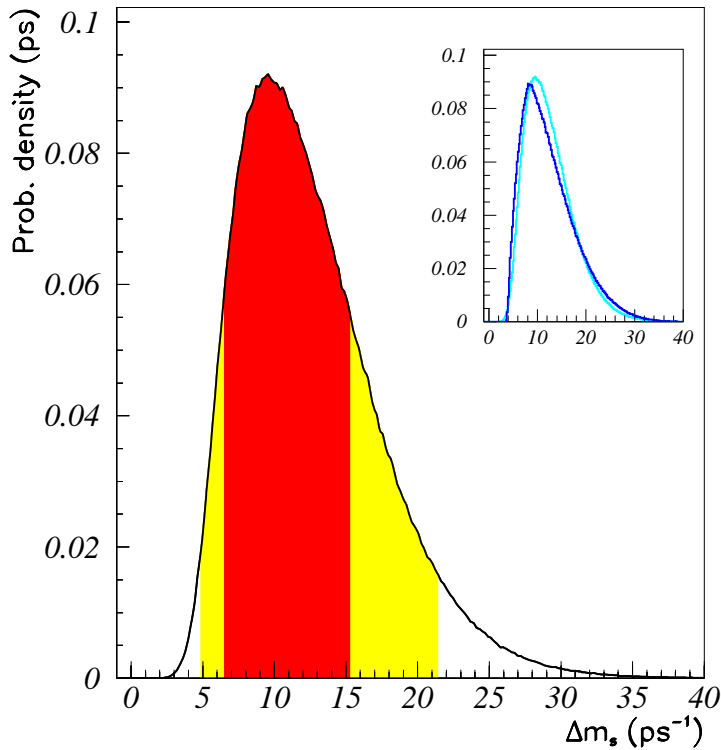,bbllx=12pt,bburx=525pt,bblly=10pt,bbury=520pt,height=10cm}}
\caption { \it{The $\Delta m_s$ probability distribution obtained with the same constraints as in Figure \ref{fig:hari_1}. 
The dark-shaded and the clear-shaded intervals correspond to 68$\%$ and 95 $\%$ confidence level regions respectively.
The inset shows the comparison between the  $\Delta m_s$ probability distributions for Gaussian distributions 
( clearer contours ) and for flat distributions ( darker countours ) }}
\label{fig:deltams}
\end{center}
\end{figure}

One of the advantage of having explicitly built up the two-dimensional probability distribution for
$\overline{\rho}$ and $\overline{\eta}$ is that the weights previously defined can be used to obtain the probability 
distribution for $\Delta m_s$. This distribution is shown in Figure \ref{fig:deltams}.  
It is expected that the value of $\Delta m_s$ is in the range between 6.5
and 15 $ps^{-1}$ at 68$\%$ C.L. The most probable value is around 10 $ps^{-1}$. At 95 $\%$ C.L. $\Delta m_s$ has to be 
smaller than 21 $ps^{-1}$. The current LEP combined lower limit at 8 $ps^{-1}$  is just exploring the one sigma 
expected region for $\Delta m_s$.
The $\Delta m_s$ distribution has been also obtained using flat distributions for the quantities in which the systematic 
error is dominant as explained at the end of Sec. \ref{sec:522}. At 68$\%$ C.L., $\Delta m_s$ 
is expected between 5 and 15 $ps^{-1}$, the most probable value is around 9 $ps^{-1}$ and at 95 $\%$ C.L. $\Delta m_s$ has to be
smaller than 22 $ps^{-1}$. The comparison between the new $\Delta m_s$ distribution with
those obtained with only Gaussian distributions is shown in the inset of Figure \ref{fig:deltams}.

\subsubsection{ Measured values for $sin2\alpha$ and $sin2\beta$ .}
\label{sec:525}

\begin{figure}
\begin{center}
{\epsfig{figure=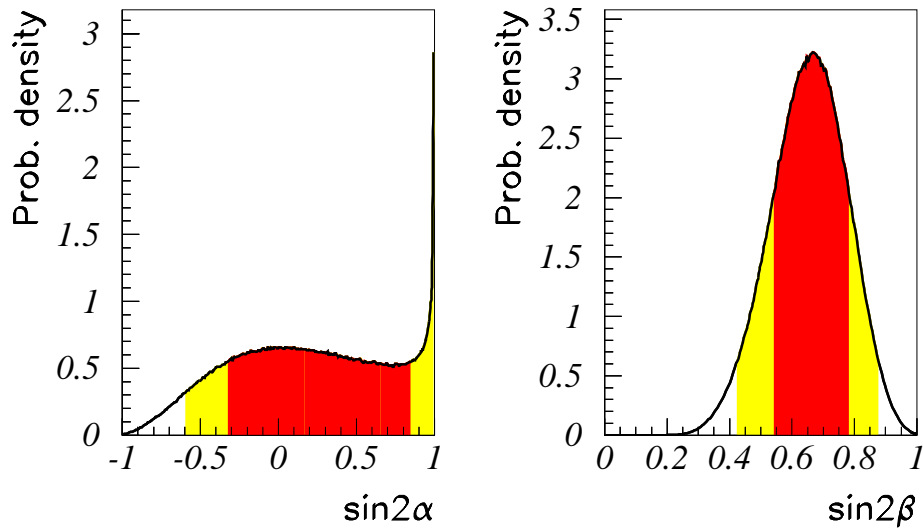,bbllx=9pt,bburx=491pt,bblly=263pt,bbury=527pt,height=7cm}}
\caption{ \it{ The sin2$\alpha$ and sin2$\beta$ distributions are shown in the left and right plots respectively. They have
been obtained using the same contraints as in Figure \ref{fig:hari_1}. 
The dark-shaded and the clear-shaded intervals correspond to 68$\%$ and 95 $\%$ confidence level regions respectively.  }}
\label{fig:seni1997}
\end{center}
\end{figure}

Present measurements of $\Delta m_d$, $\left | V_{ub} \right |$,
$ \mid \epsilon_K \mid $ and the limit on $\Delta m_s$ can be used also to define the allowed region in the plane 
($sin2\alpha, sin2\beta$), where $\alpha$ and $\beta$ are respectively equal to the 
angles $\widehat{CAB}$ and $\widehat{ABC}$ of the unitarity triangle. The values 
of $sin2\alpha$ and $sin2\beta$ are expected to be measured at future B-factories
from the studies of the decays $\mbox{B}^0_d \rightarrow \pi^+ \pi^-$ and
$\mbox{B}^0_d \rightarrow J/\Psi \mbox{K}^0_s$, respectively.
With present measurements the following values have been obtained (Figure \ref{fig:seni1997}):
\begin{equation}
sin 2\alpha~ >~-0.60 ~ at ~ 95 \% C.L , ~~~~ sin 2\beta~=~ 0.67 ^{+0.12}_{-0.13}  
\end{equation}

The accuracy on $sin 2\beta$ is already at the level expected in the year 2000
at B-factories. The value of $sin 2\alpha$ is more uncertain because it depends 
a lot on the measurement of $\overline{\rho}$; for the fitted values of $\overline{\rho}$ and $\overline{\eta}$,
the error on $\sigma(sin 2\alpha)$ is typically $\simeq 5~\sigma(\overline{\rho})$. 
Using flat distributions for the quantities in which the systematic error is dominant, as explained in Sec. \ref{sec:522}, 
the new value for $sin 2\beta$ is equal to $0.82 ^{+0.05}_{-0.22}$.

\section{ The perspectives }
\label{sec:6}

\subsection{Improvements on the determination of non-perturbative QCD parameters}
\label{sec:61}

\subsubsection { The $f_B$ parameter }
\label{sec:611}

It has been shown in section ~\ref{sec:523}  that the accuracy on the measurement of $\overline{\rho}$ can be improved
if $f_{B_d} \sqrt{{B_{B_d}}}$ can be obtained with an uncertainty smaller than 30 MeV.
 
\underline {Direct measurement of $f_{B}$.}
A priori, $f_{B}$ can be measured using the decay  $B^+ \rightarrow \tau^{+} \nu_{\tau}$ which
proceeds through the annihilation of the $\overline{b}$ and $u$ quarks. The 
corresponding branching fraction for this process is proportional to the 
product $f^2_{B} \times \left | V_{ub} \right |^2$. Thus, following this procedure, $f_{B}$ is not directely measured. 
Furthermore the Standard Model predicts a branching fraction
of the order of $7 \times 10^{-5}$ which is very small as compared to possible sources of background. 
Present limits from LEP and CLEO experiments are at the level  of  $5 \times 10^{-4}$  \cite{ref:bptau_lep}
and  $2.2 \times 10^{-3}$ at $90 \%$ C.L. \cite{ref:bptau_cleo} respectively. 
From the analysis point of view LEP is the most favourable place to perform this measurement but the statistics are too low.
At CLEO and future B-factories this measurement seems to be very difficult and a 10$\%$ precision seems to be 
unrealistic. \\

\underline {$f_B$ from lattice QCD calculations: the use of $f_{D_{s}}$ to evaluate $f_B$}. 
The study of the decay $D^+_s \rightarrow \tau^{+} \nu_{\tau}$ or  $D^+_s \rightarrow \mu^{+} \nu_{\mu}$ has two
clear advantages with respect to the $B^+ \rightarrow \tau^{+} \nu_{\tau}$ channel. The branching fraction is of the order of
5 $\%$, for the first channel, and $f_{D_{s}}$ can be directely obtained, since the value of the CKM element 
$\left | V_{cs} \right |$ is well known.\\
From the present measurements of $D^+_s \rightarrow \tau^{+} \nu_{\tau}$ and $D^+_s \rightarrow \mu^{+} \nu_{\mu}$ decays, 
$f_{D_{s}}$ is derived to be  \cite{ref:dstau} :

\begin{equation}
f_{D_{s}} = 255 \pm 20 \pm 31 MeV
\label{eq:eqdstau}
\end{equation}
The method proposed in the following, to evaluate $f_B$, consists in using this 
 measurement and the extrapolation from the D to the B sector, as predicted by lattice QCD \cite{ref:soni}.
%in fact, once $f_{D_{s}}$ is precisely measured, the results on ratio of the others decay constants, which are known with a
%precision of $5\%$, can be used to reduce the errors on the value of $f_B$.
At present, the ratio $f_{D_{s}} / f_{D^+}$ has also to be taken from the theory
and the value given in \cite{ref:dconstant} has been used:
$$
\frac{f_{D_{s}}}{ f_{D^+}} = 1.10 \pm 0.05 \pm 0.10
$$
where the last error is an estimate of uncertainties from the quenching approximation.
The ratio  $f_B / f_D $ is then obtained using the
expected variation, from lattice QCD, of $f \times \sqrt {M}$ as a function of 1/M and requiring that 
the prediction co\"\i ncides with the measured decay constant in the D region.
From the dispersion of present predictions, it has been assumed that this extrapolation gives
an additional 10$\%$ relative error. 
It results that:
\begin{equation}
 f_{B_d} = 178 \pm 26(exp.) \pm 18 ( \frac{f_{D_{s}}}{ f_{D^+}} ) \pm 18 (\frac{f_{B}}{ f_{D^+}} ) MeV
\label{eq:fb_exp}
\end{equation}

The value obtained for $ f_{B_d} $, in this approach is well compatible with the absolute prediction from
lattice QCD (eq. \ref{eq:fb}) and also with the value favoured by the measurements of $ \mid \epsilon_K \mid $, $\Delta m_d$, 
$\frac{\left | V_{ub} \right |}{\left | V_{cb} \right |}$ and the limit on $\Delta m_s$, corresponding to eq. (\ref{eq:fb_mes})
\begin{equation}
 f_{B_d} = 178 ^{+26}_{-34} ~~MeV
\label{eq:fb_fit}
\end{equation}

This approach seems promising because accurate experimental measurements can be obtained in future. In the coming
years, improvements are expected from CLEO and from LEP on the absolute decay rate $D^+_s \rightarrow \tau^+ \nu_{\tau}$
( $D^+_s \rightarrow \mu^{+} \nu_{\mu}$ ).
A spectacular improvement is expected from a future Tau-Charm factory where $f_{D_s}$ and $f_{D^+}$ can be measured with
a relative accuracy of $1.5 \%$. These two measurements can provide very strict constraints on results from lattice QCD. 
In particular they can validate unquenched evaluations when going from bound states made with an heavy quark and a strange
or a non-strange anti-quark. 
On the other hands a theoretical effort has to be done to further reduce the error from the extrapolation between
the D and B mass regions.

\subsubsection { $B_{B_d}$ and $B_K$ parameters.}
The bag parameters $B_K$ and $B_{B_d}$ cannot be measured experimentally. As it has been shown in Sections \ref{sec:42},
\ref{sec:43} the precision on these parameters is limited by the error related to the quenched approximation. New and more precise
results expected in the next future and a better control of the quenched approximation should allow to reach a precision of
5 $\%$ on $B_K$ and $B_{B_d}$. 

\label{sec:612}

%These parameters cannot be experimentally measured.
%For the $B_B$ parameter the precision previously considered of $\pm0.09$ comes from a conservative balance between the lattice QCD
%and the sum rules calculations. The precision from lattice QCD calculation is twiced better. The same holds for the $B_K$ 
%parameter. In this case the precision coming from the lattice QCD calculation id $\pm 0.03$ to be compared to the conservative
%uncertainty of $\pm 0.15$ which has been assumed to take into account the calculation done in the 1/N approach.

\subsection { Improvements on the determination of the $\Delta m_s$ frequency}
\label{sec:62}

The best limit on $\Delta m_s$ ( $\Delta m_s > 8.0 ps^{-1} ~at~95 \% C.L.$ ) is coming from the combination of the latest
analyses from ALEPH, DELPHI and OPAL collaborations at LEP. 
Improvements are expected by adding new analyses, from the refinement of present ones and from new results from the SLD experiment.
From a study, done in DELPHI, the final LEP sensitivity \footnote{ In the amplitude approach described in Section \ref{sec:522}, 
it is possible to compute the exclusion 
probability ${\cal P}_{limit}$ of being able to set a limit for a given value of $\Delta m_s$, with the studied channel. The
sensitivity is the value of  $\Delta m_s$ corresponding to ${\cal P}_{limit} = 0.5$.}
for $\Delta m_s$ will be around 12.5 ps$^{-1}$ \cite{ref:highl} which is
above the expected most probable value of 10 $ps^{-1}$ as shown in Figure \ref{fig:deltams}.
It has been also demonstrated \cite{ref:highl} that this result is
mainly limited by the available statistics collected by the  experiments. A gain of a factor four in statistics would
allow to have a sensitivity up to $\Delta m_s=16$ ps$^{-1}$. In this case, for a value of $\Delta m_s=10$ ps$^{-1}$, 
which corresponds to the most probable value, at present, a measurement, with a precision better than 15 $\%$ could also 
be performed \cite{ref:highl}.

\subsection { Improvements on $A$, $m_t$ and $\frac{\left | V_{ub} \right |}{\left | V_{cb} \right |}$ }
\label{sec:63}

As shown in Section \ref{sec:3} the measurement of  $ \mid V_{cb} \mid $ comes from the average of two methods. The first one uses
the values of the inclusive lifetime and of the inclusive semileptonic
branching fraction of B hadrons. Its precision is limited 
by the accuracy on the theoretical parameter $\gamma_c$. An improvement on its determination is expected \cite{ref:varenna},\cite{ref:bigivbc}. 
The second method uses the measurement of $Br (\mbox{B} \rightarrow \mbox{D}^* \ell \nu$) in the framework of HQET. The precision of this measurement
is limited by experimental errors. On statistical errors, improvements are expected mainly from the CLEO experiment.
The main systematic uncertainties come from the errors on the $B_d^0$ lifetime, on the  $B_d^0$
production rate in jets and on the $\mbox{D}^{**}$ production rate in B hadron semileptonic decays. 
These parameters can be more precisely measured at LEP in the near future. 
All these improvements could allow to reach a precision of $\pm 0.001$ on $ \mid V_{cb} \mid $ which 
corresponds to an uncertainty of $ \pm 0.025 $ on $A$.\\
Theoretical improvements are also expected to extract the ratio $\frac{\left | V_{ub} \right |}{\left | V_{cb} \right |}$ 
\cite{ref:bigivbc}. 
A possible error of $\pm 0.01$ will be considered in the perspectives. \\
In addition the final analyses of CDF and D0 at Fermilab should allow to reach a precision of $\pm 5 GeV$ on the top mass.

\subsection { Different scenarii.}
\label{sec:64}

Three scenarii are presented in this section. It has to be noticed that the central values obtained in each case
for $\overline{\rho}$, $sin 2 \alpha$ or $f_B$ are rather arbitrary because the informations on these quantities
are obtained with the present central values and reduced uncertainties on the different parameters entering into
the measurements. Only quoted uncertainties on these quantities are meaningful.

\subsubsection {``Year 2000" scenario. } 

\begin{figure}
\begin{center}
{\epsfig{figure=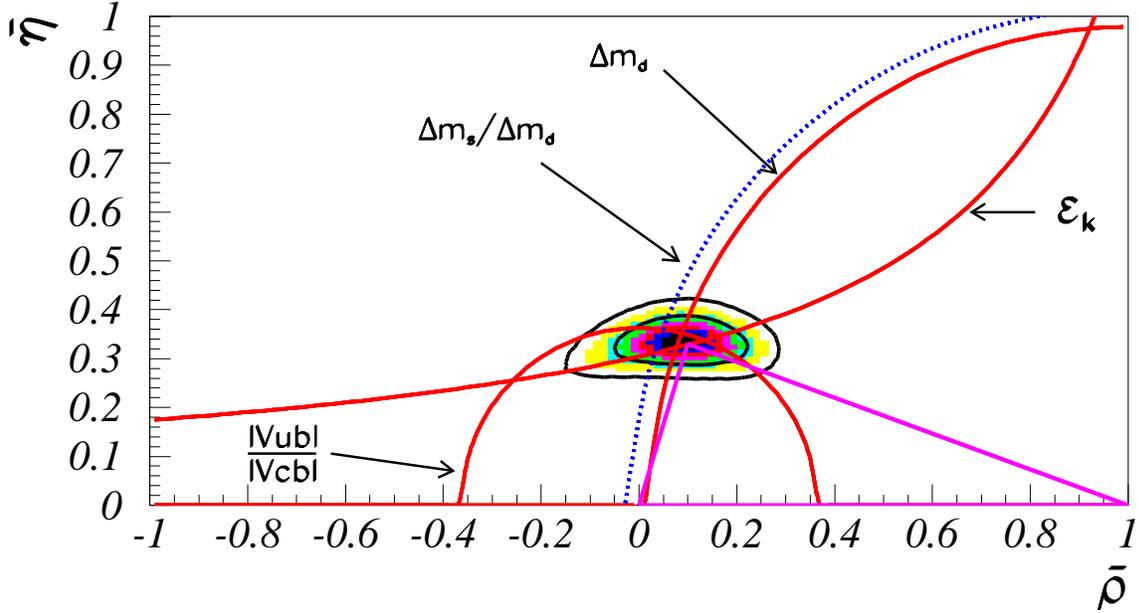,bbllx=30pt,bburx=489pt,bblly=6pt,bbury=251pt,height=8cm}}
\caption{ \it{ The $\overline{\rho}$-$\overline{\eta}$ allowed region in the ``Year 2000" scenario. 
Contours at 68$\%$ and 95 $\%$ are indicated.}}
\label{fig:anno2000}
\end{center}
\end{figure}

\begin{figure}
\begin{center}
{\epsfig{figure=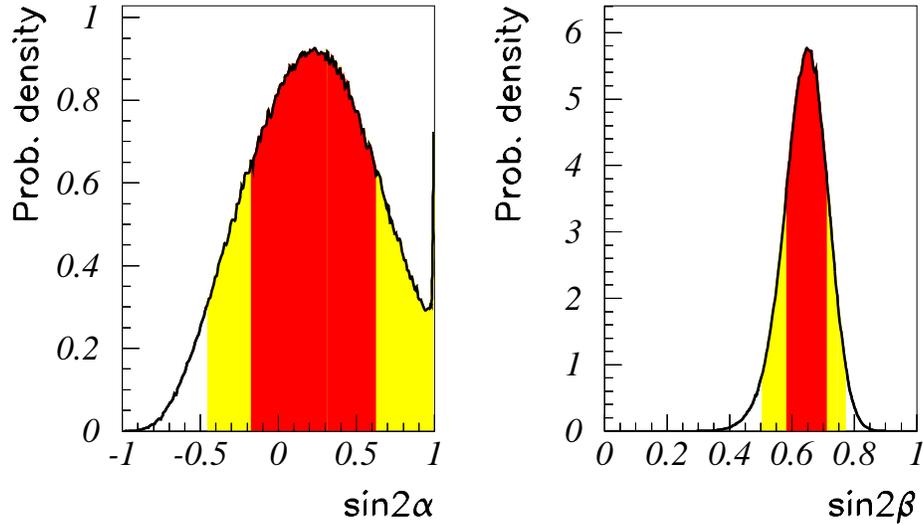,bbllx=9pt,bburx=491pt,bblly=263pt,bbury=527pt,height=7cm}}
\caption{ \it{ The $sin 2 \alpha$ and $sin 2 \beta$ distributions, shown in the left and right plots respectively, have been
obtained using the contraints corresponding to the values of the parameters expected in the ``Year 2000" scenario. 
The dark-shaded and the clear-shaded intervals correspond to 68$\%$ and 95 $\%$ confidence level regions respectively. }}
\label{fig:seni2000}
\end{center}
\end{figure}

The central values and uncertainties of the parameters, used for this scenario, 
are given in the third column of Table \ref{tab:a}. This scenario, named ``Year 2000", corresponds to the possible 
situation at the threshold of the year 2000, before the start up of B-factories.\\
It includes the "possible" latest CLEO and LEP results. A sensitivity at 12.5 ps$^{-1}$ on $\Delta m_s$ is considered and, 
in addition, better estimates for the $f_B$, $B_B$, $\xi$ and $B_K$
parameters on which uncertainties have been reduced by a factor of about two. The error used for $f_B$ comes from the 
current determination of $f_{D_{s}}$ whereas the central value is kept the same as in eq. (\ref{eq:fb}).
The improvements on $A$, $m_t$ and  $\frac{\left | V_{ub} \right |}{\left | V_{cb} \right |}$,
described in previous paragraphs, have been included. The results are:
\begin{equation}
\overline{\rho} = 0.09 ^{+0.07}_{-0.11} ~~ ; ~~\overline{\eta} = 0.335 ^{+0.034}_{-0.033} 
\label{eq:ciao}
\end{equation}

The new allowed region for the $\overline{\rho}$ and $\overline{\eta}$ parameters is shown in Figure \ref{fig:anno2000}.
It corresponds to:
\begin{equation}
sin 2 \alpha~ =~0.24 \pm 0.41  ; ~~~~ sin 2 \beta~=~ 0.65 ^{+0.06}_{-0.07} 
\end{equation}
The $sin 2 \alpha$, $sin 2 \beta$ distributions are shown in Figure \ref{fig:seni2000}.\\
\vskip 0.2truecm
Two scenarios are presented in the following. The "$\tau$-Charm" scenario which 
considers the possibility of having a $\tau$-Charm factory operating at the beginning 
of the next millenium, allowing an accurate control of non-perturbative QCD parametrs.
The CP-phases scenario which shows the importance of the direct measurement of 
$sin 2 \alpha$ and $sin 2 \beta$. 
Several experimental facilities have been approved for the latter purpose
whereas none  $\tau$-charm factory is foreseen yet. The aim of separately showing 
the two scenarios is to stress their complementary and to underline, at least once, 
the importance of having a $\tau$-Charm factory project soon operating.

\subsubsection{ "$\tau$-Charm" scenario. }

\begin{figure}
\begin{center}
{\epsfig{figure=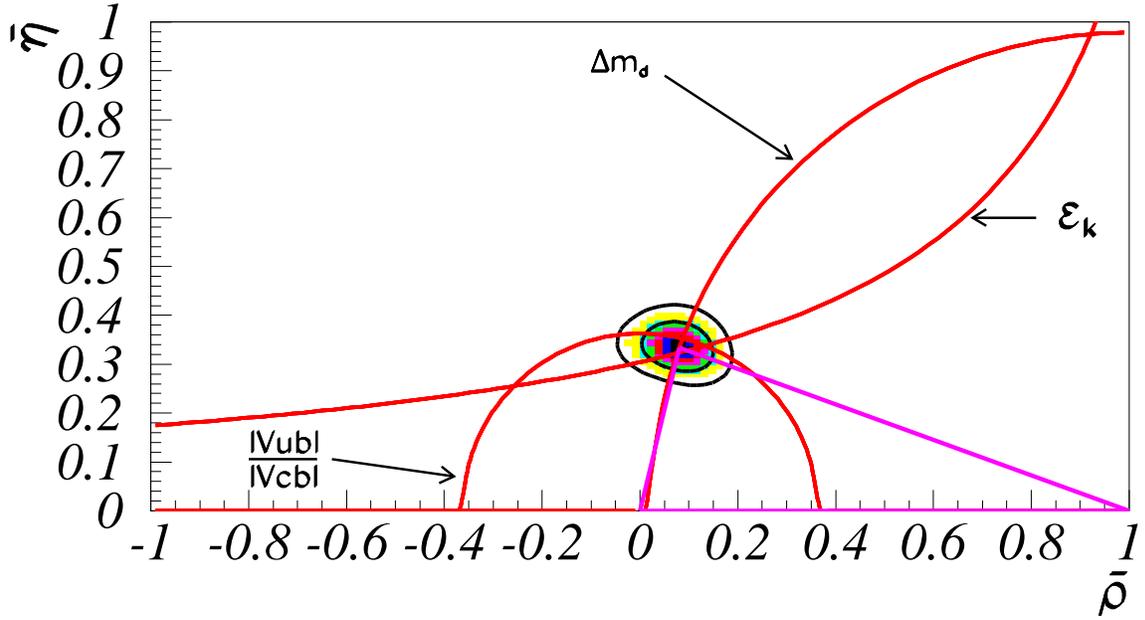,bbllx=30pt,bburx=489pt,bblly=6pt,bbury=251pt,height=8cm}}
\caption{ \it{ The $\overline{\rho}$-$\overline{\eta}$ allowed region in the ``$\tau$-Charm" scenario.
Contours at 68$\%$ and 95 $\%$ C.L. are indicated.}}
\label{fig:taucharm}
\end{center}
\end{figure}

This scenario considers the possibility of having a $\tau$-Charm factory operating at the beginning of the next millenium.\\ 
In the previous scenario it has been shown that the $\overline{\rho}$ parameter is still known 
three times less precisely than the $\overline{\eta}$
parameter. This is mainly due to the uncertainty on $f_B$. At a $\tau$-Charm factory $f_{D_{s}}$ and $f_{D^+}$ can be
measured with a precision at the per cent level. Following the approach described in \ref{sec:611}, $f_B$ may be 
determined with a precision better than $5 \%$. To reach this goal a joint effort, between experimentalists and theorists
 working on lattice QCD, is needed.

\begin{figure}
\begin{center}
{\epsfig{figure=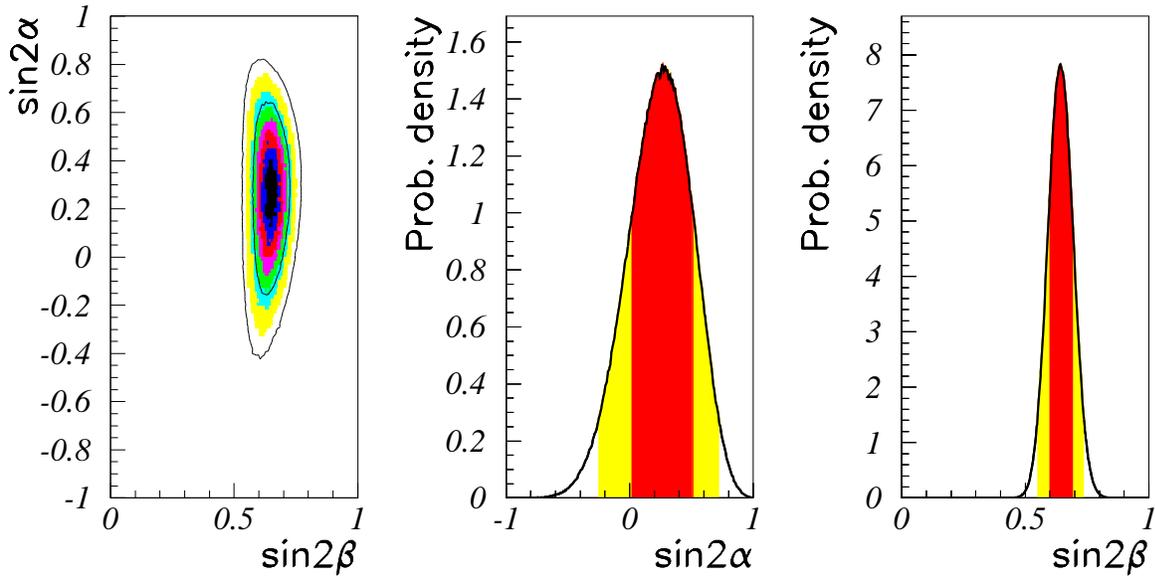,bbllx=70pt,bburx=660pt,bblly=323pt,bbury=669pt,height=8cm}}
\caption{ \it{The $sin 2 \alpha$ and $sin 2 \beta$ distributions have been
obtained using the contraints corresponding to the values of the parameters expected in the ``$\tau$-Charm" scenario.
 The dark-shaded and the clear-shaded intervals correspond, respectively, to 68$\%$ and 95 $\%$ confidence level regions.}}
\label{fig:senitaucharm}
\end{center}
\end{figure}

Assuming 2.5 $\%$ accuracy on the quantity $f_{B_d} \sqrt{B_{B_d}}$ ($f_{B_d} \sqrt{B_{B_d}}~=~200 \pm 5~MeV$), 
the expected accuracy on the $\overline{\rho}$ and $\overline{\eta}$ parameters comes out to be:
\begin{equation}
\overline{\rho} = 0.080 \pm 0.048 ~~ ; ~~\overline{\eta} = 0.334 \pm 0.034 
\label{eq:res_taucharm}
\end{equation}
and the new allowed region is shown in Figure \ref{fig:taucharm}.
The information coming from $\Delta m_s$ result is not used, the reason being that the measurement of $f_{B_d} \sqrt{B_{B_d}}$ 
and of $\frac{\Delta m_d}{\Delta m_s}$ give the same type of constraint. In practice the constraint at $\pm 5$ MeV on
$f_{B_d} \sqrt{B_{B_d}}$ has the same effect as a measurement of $\Delta m_s = 12.8 \pm 1.8~ps^{-1}$.\\
The corresponding values for $sin 2 \alpha$ and $sin 2 \beta$ are:
\begin{equation}
sin 2 \alpha~ =~0.23 \pm 0.25  ; ~~~~ sin 2 \beta~=~ 0.65 \pm 0.04,
\end{equation}
and their expected distributions are shown in Figure \ref{fig:senitaucharm}.\\
The inclusion of the $\Delta m_s$ constraint slightly improves the errors on  the $\overline{\rho}$ and $\overline{\eta}$ 
parameters.

\begin{table}
\begin{center}
\begin{tabular}{|c|c|c|c|c|}
\hline 
Scenarii                   &  $\overline{\rho}$  & $\overline{\eta}$  & $\Delta m_s$ (ps$^{-1}$) & $f_B \sqrt{B_{B}}(MeV)$  \\
\hline 
At-present                 & $0.10^{+0.13}_{-0.38}$ & $0.33^{+0.06}_{-0.09}$ & $10^{+5.0}_{-3.5}$           &      -        \\ 
At-present + $\Delta m_s$  & $0.11^{+0.13}_{-0.25}$ & $0.33 \pm 0.06$    & $[>8~ps^{-1}~at~95\%C.L.]$ &  $208^{+30}_{-40}$ \\ \hline
Year-2000           & $0.09^{+0.07}_{-0.11}$ & $0.335 ^{+0.034}_{-0.035}$ & $[sensitivity~at~12.5~ps^{-1}]$ &   $\pm 15$    \\ \hline
$\tau$-Charm               & $0.080 \pm 0.048$ & $0.335 \pm 0.034$   &     $12.8 \pm 1.8 ~ps^{-1}$      & $[ \pm 5 ]$   \\ \hline
CP-phases                  & $0.121^{+0.029}_{-0.026}$  & $0.331 \pm 0.026$   & $[sensitivity~at~12.5~ps^{-1}]$ &  $\pm 9$ \\ \hline
\end{tabular}
\caption[]{ \it {Results on $\overline{\rho}$-$\overline{\eta}$,
 $\Delta m_s$ and $f_{B_d} \sqrt{B_{B_d}}$ in the different scenarii. Apart from the present
measurements, the central values of these quantities -with the exception of the one for $\overline{\eta}$-
are not meaningful, only expected errors have to be retained. The results on $\Delta m_s$ and on $f_B \sqrt{B_{B}}$ are given in
squared brackets when they correspond to values given as constraints and which are not improved at the end of the fitting
procedure.} }
\label{tab:scenario1}
\end{center}
\end{table}

\subsubsection{``CP-phases" scenario.}

\begin{figure}
\begin{center}
{\epsfig{figure=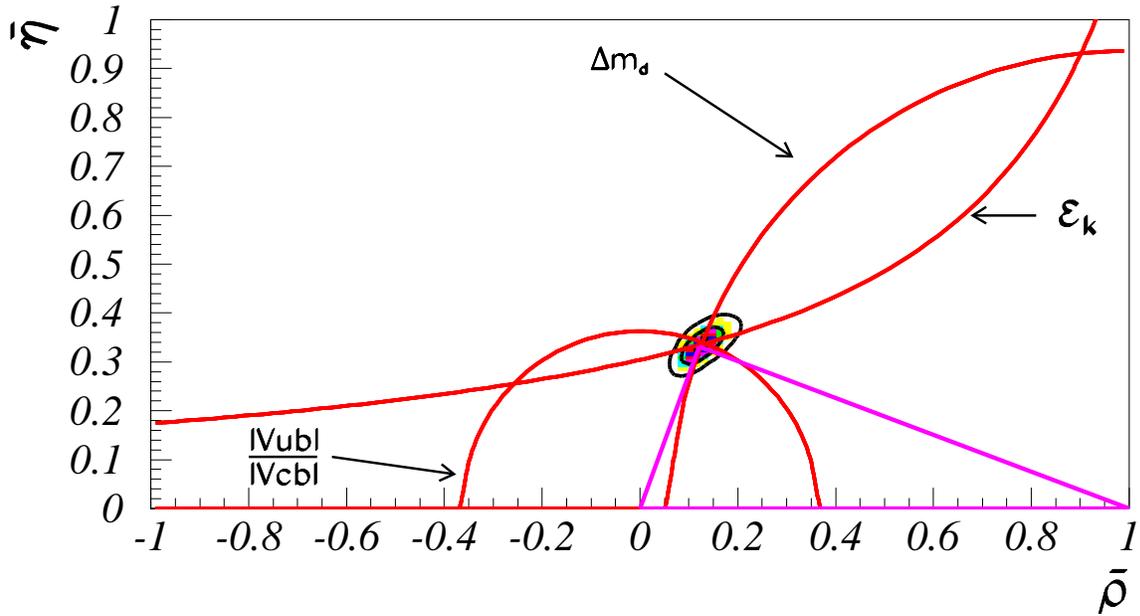,bbllx=30pt,bburx=489pt,bblly=6pt,bbury=251pt,height=8cm}}
\caption{ \it{  The $\overline{\rho}$-$\overline{\eta}$ allowed region in the ``CP-phases" scenario.
Contours at 68$\%$ and 95 $\%$ C.L. are indicated.}}
\label{fig:usinebb}
\end{center}
\end{figure}

\begin{figure}
\begin{center}
{\epsfig{figure=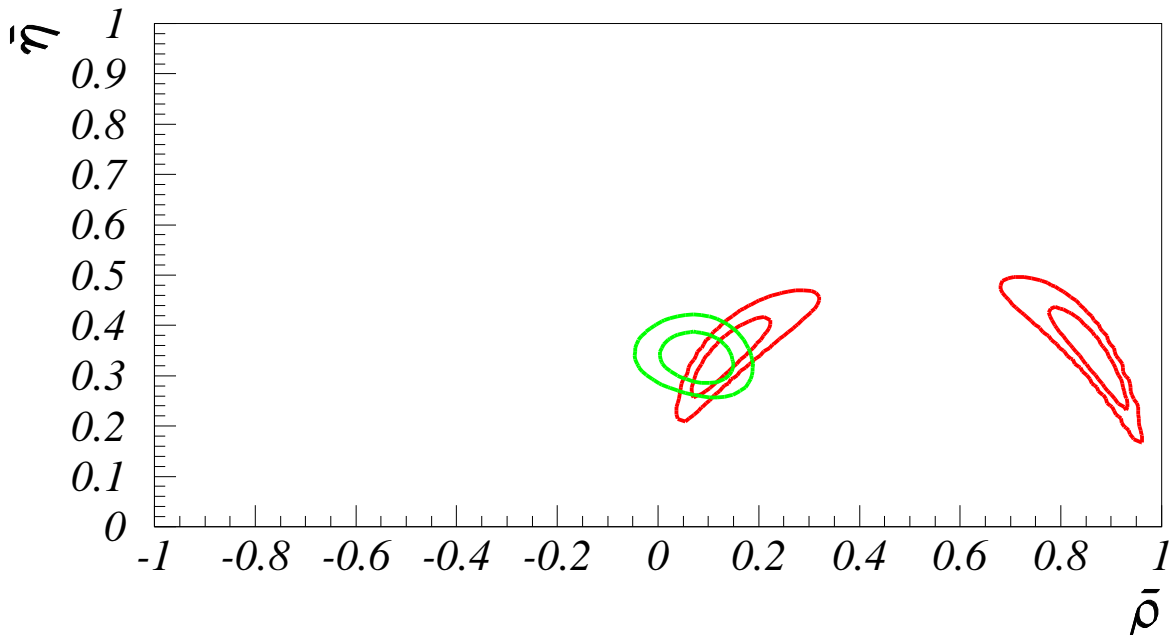,bbllx=21pt,bburx=494pt,bblly=10pt,bbury=248pt,height=8cm}}
\caption{ \it{ The contours at 68$\%$ and 95 $\%$ C.L. in the $\overline{\rho}$-$\overline{\eta}$ plane obtained
 in the ``$\tau$-Charm" scenario ( clearer contours ) are compared with those 
coming from the measurement of the quantities $sin 2 \alpha$ and $sin 2 \beta$, alone, with $\pm$ 0.10 accuracy at future
B-factories ( darker contours ).}}
\label{fig:usinebb_taucharm}
\end{center}
\end{figure}

In this scenario, the direct measurement of $sin 2 \alpha$ and $sin 2 \beta$ is considered.\\
Several experiments have been approved for this purpose. HERA-B at DESY and the Fermilab experiments ( the upgrades of
D0 and CDF experiments ) are mainly sensitive to $sin 2\beta$ via the reconstruction of the decay 
$\mbox{B} \rightarrow J/ \psi \mbox{K}_{s}^{0}$. As shown in Section \ref{sec:525} the current precision on $sin 2\beta$,
obtained from indirect measurements in the framework of the Standard Model and of the C.K.M. matrix, is $\pm 0.13$  and it is 
expected to improve down to 
$\pm 0.07$ in the year 2000. The HERA-B experiment foresees to have a precision on $sin 2 \beta$ of the order of  $\pm 0.20$
%which is of any interest. At the contrary the 
and an accuracy of $\pm 0.10$ is expected at Fermilab experiments.\\
The $e^+ e^-$ asymmetric B-factories ( BaBar at SLAC and BELLE at KEK ) can measure both $sin 2\beta$ and $sin 2\alpha$, the 
latter via the reconstruction of the decay $\mbox{B} \rightarrow \pi^+ \pi^-$.\\
Assuming that these two quantities are measured with a precision of:
$$
     \delta(sin 2 \beta) =  \pm 0.10    ~~~~ ; ~~~~ \delta(sin 2 \alpha) =  \pm 0.10 
$$
and considering that the other parameters, listed in Table \ref{tab:a}, are known with an accuracy corresponding to the 
``Year-2000" scenario, the expected precisions on $\overline{\rho}$ and $\overline{\eta}$ are then:
\begin{equation}
\overline{\rho} = 0.121 ^{+0.029}_{-0.026} ~~ ; ~~\overline{\eta} = 0.331  \pm 0.026
\label{eq:ciao1}
\end{equation}
This corresponds to an equivalent precision of $\pm 9 MeV $ on the $f_{B_d}$ parameter.\\
The new $\overline{\rho}$-$\overline{\eta}$ allowed region is shown in Figure \ref{fig:usinebb}. 
It should be reminded that B-factories will contribute to reduce the $\overline{\rho}-\overline{\eta}$ 
region by performing also additional measurements on $\frac{\left | V_{ub} \right |}{\left | V_{cb} \right |}$ 
and on $\left | V_{cb} \right | $.\\
It will be of interest to compare the  $\overline{\rho}$-$\overline{\eta}$ allowed regions in the two following different 
approaches :

\begin{itemize}
\item
the precise measurements of the sides of the unitary triangle, which requires an accurate control of non-perturbative QCD 
parameters (``Year-2000" and "$\tau$-Charm" scenarii) corresponding to a determination of $f_{B_d}$ with an error of 
10 MeV or better,
\item 
the precise measurements of the angles (``CP-phases" scenario).
\end{itemize}
The two selected $\overline{\rho}$-$\overline{\eta}$ allowed regions are shown in Figure \ref{fig:usinebb_taucharm}. 
%It is then clear from it than the $\pm 5 MeV $ error on $f_B$ is necessary to select region of similar sizes in terms 
%of $\rho$ and $\eta$ parameters in the two approaches.

A summary of the results expected on  $\overline{\rho}$-$\overline{\eta}$,
 $\Delta m_s$ and $f_{B_d} \sqrt{B_{B_d}}$ in the different scenarii is
given in Table \ref{tab:scenario1}. Table \ref{tab:scenario2} gives the summary for the $sin 2\alpha$ and $sin 2 \beta$ variables.

\begin{table}
\begin{center}
\begin{tabular}{|c|c|c|}
\hline
Scenarii           &    $sin 2 \alpha$          &             $sin 2 \beta$        \\
\hline
At-present                     &     $>-0.60~at~ 95 \% C.L$ &         $0.67^{+0.12}_{-0.13}$    \\ \hline
Year-2000                      &      $0.24 \pm 0.41 $      &         $0.65 ^{+0.06}_{-0.07}$   \\ \hline
$\tau$-Charm                   &      $0.23 \pm 0.25 $      &         $0.65 \pm 0.04$           \\ \hline 
CP-phases                      &      $[ \pm 0.10 ]  $      &         $ [ \pm 0.10 ]$           \\ \hline 
\end{tabular}
\caption[]{ \it {Results for $sin 2\alpha$ and $sin 2 \beta$ in the different scenarii. For the CP-phases scenario the errors refer to
the direct measurement of the angles.}}
\label{tab:scenario2}
\end{center}
\end{table}

\section { Possible effects from supersymmetric particles.}
\label{sec:8}
In this section are discussed the additional contributions to $\Delta m_d$ and to $ \mid \epsilon_K \mid $ expected to come from
the presence of new physics beyond the Standard Model. The effect of new physics can be parametrized by introducing an extra
parameter $\Delta$. This formulation is valid in any extension of the Standard Model in which the flavour changes are controlled 
by the $V_{CKM}$ matrix. \\
In case, for instance, of $\Delta m_d$ the expression (\ref{eq:deltamd1}) becomes
\begin{equation}
\Delta m_d~=~\frac{g^4}{192 m_{W}^{2} \pi^2} \mid V_{tb} \mid^2 \mid V_{td} \mid^2
m_{B_d} f^2_{B_d} B_{B_d} \eta_B \Delta 
\label{eq:dmd_susy}
\end{equation}
The same quantity $\Delta$ appears in the expression of $ \mid \epsilon_K \mid $.\\  
The $\Delta$ parameter can be written
\begin{equation}
\Delta  =  x_t ( F(x_t) + \Delta_{New Physics} )
\label{eq:delta}
\end{equation}
The Standard Model predicts : $\Delta=2.55 \pm 0.15 $, where the error takes into account the uncertainty on the top mass.
The present data can be then fitted using the new expressions for $\Delta m_d$ and $ \mid \epsilon_K \mid$
fitting $\Delta$ together with the parameters $\overline{\rho}$ and $\overline{\eta}$, the result is :
$$
\overline{\rho} = 0.20 ^{+0.19}_{-0.38} ~~ ; ~~ \overline{\eta} = 0.29 ^{+0.13}_{-0.15}    \\
$$
\begin{equation}
\Delta = 3.2^{+5.9}_{-1.9}   
\label{eq:fitsusy}
\end{equation}

\begin{table}
\begin{center}
\begin{tabular}{|c|c|c|c|c|}
\hline
 Scenarii                   &  $\Delta$     & $m_{SUSY}$ , tan $\beta$ = 1 &  tan $\beta$=1.5 &  tan $\beta$ = 5    \\ \hline
 At-present                 & $3.2^{+5.9}_{-1.9}$  &           $>55$              & no limit         &    no limit  \\ \hline
 Year-2000                  & $\pm $1.0            &           $>120$             &  $>85$           &     $>60$    \\ \hline
 $\tau$-Charm               & $\pm $0.7            &           $>140$             &  $>105$          &     $>75$    \\ \hline
 CP-phases                  & $\pm $0.6            &           $>150$                & $>115$        &     $>80$    \\ \hline
\end{tabular}
\caption[]{ \it {Results on the $\Delta$ parameter in the different scenarii. The limits on $m_{SUSY}$ are given 
in $GeV/c^{2}$ at 95 $\%$ C.L.}}
\label{tab:susy1}
\end{center}
\end{table}

The same fit has been performed for the scenarii presented in the previous section and
results have been summarized in Table \ref{tab:susy1}. 
Since the central values of all parameters entering into the fit have been fixed at the presently measured ones,
only the expected error on $\Delta$ is given. \\
To see the possible implications of such a precision on the $\Delta$ parameter a particular example is discussed in the following,
in the framework of the MSSM extension of the Standard Model, stressing the constraints that can be obtained essentially on the 
masses of the lightest supersymmeric particles. 
The theoretical framework is discussed in \cite{ref:susy}. In this framework light stop-right ($\tilde {t}$) and
higgsinos are considered assuming that $tan ^2 \beta$ is lower than $m_t/m_b$ and that the stop-left and the gauginos are heavy.  
In this scenario $\Delta$ can be written :
\begin{equation}
\Delta_{New Physics} = ( \Delta_H + \Delta_{SUSY} ) 
\label{eq:dmd_susy_zwi}
\end{equation}
$\Delta_H$ and $\Delta_{SUSY}$ denote the contributions to the box diagram from charged Higgs boson and from R-odd supersymmetric
particles respectively; for the latter, contributions of the stop-right ($\tilde {t}$) and of the charged higgsino 
($\tilde{\chi}$) are considered.
The expression for these quantities are given in \cite{ref:susy}. In the region of masses between 60 and 150 GeV/c$^2$, 
$\Delta_{SUSY}$ dominates over $\Delta_{H}$ because the former depends on $(m_{\tilde {t}} + m_{\tilde{\chi}})/2$ and is enhanced 
by the presence of the factor $(m^2_{t}/m^2_{SUSY})$ and, due to the term $1/sin^4(\beta$), is rapidly growing for 
$tan \beta \simeq 1$.

\begin{figure}
\begin{center}
{\epsfig{figure=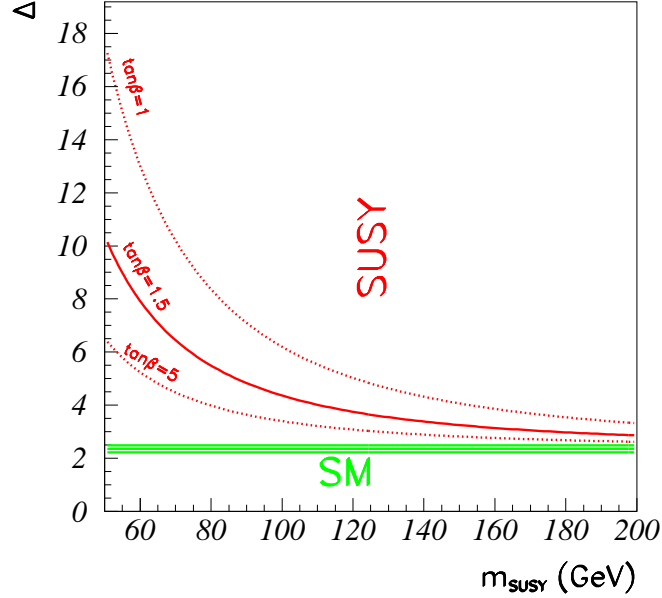,bbllx=80pt,bburx=501pt,bblly=4pt,bbury=390pt,height=8cm}}
\caption{ \it{ The curves represent $\Delta$ as a function of $m_{SUSY}$ for different values of $tan \beta$. The
horizontal lines correspond to the value of $\Delta$ expected in the Standard Model with the uncertainty due to the top mass.}}
\label{fig:susyfig}
\end{center}
\end{figure}

\begin{figure}
\begin{center}
{\epsfig{figure=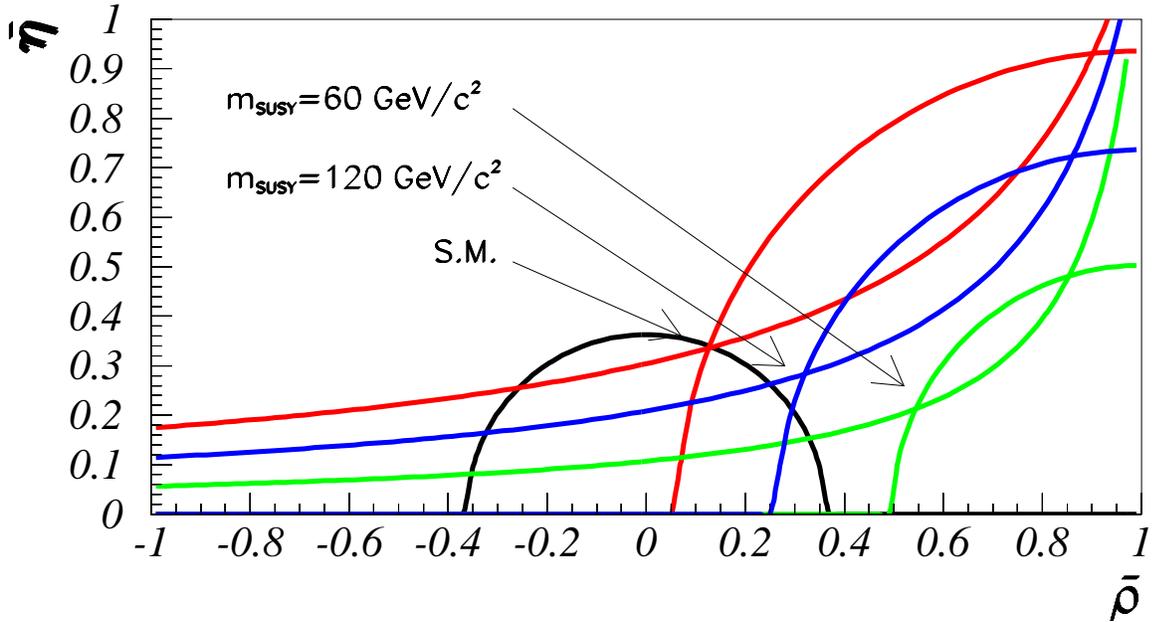,bbllx=40pt,bburx=494pt,bblly=4pt,bbury=246pt,height=8cm}}
\caption{ \it{ The compatibility of the constraints on $\Delta m_d$, $ \mid \epsilon_K \mid $ and 
$\frac{\left | V_{ub}\right |}{\left | V_{cb}\right |}$ is shown in three different scenarii. The constraint on 
$\frac{\left | V_{ub}\right |}{\left | V_{cb}\right |}$ does not change while from left to right the constraints on
$\Delta m_d$ and $ \mid \epsilon_K \mid $ change from the Standard Model scenario to two SUSY scenarii with $m_{SUSY}$ = 
120 GeV/c$^2$ and 60 GeV/c$^2$ and tan$\beta$ = 1.5.}}
\label{fig:compa}
\end{center}
\end{figure}

In the following the term $\Delta_{H}$ has been neglected and the stop-right and the charged higgsino are supposed to have the same
mass (generically indicated as $m_{SUSY}$ in the following). The quantity $\Delta$ is then defined as:

\begin{equation}
\Delta = x_t (F(x_t) + \Delta_{SUSY}(x_{t_{\tilde{\chi}}},x_{\tilde{t}_{\tilde{\chi}}},\beta))
\label{eq:delta_zwi}
\end{equation} 
\noindent
where $x_t$ = $(m^2_{t}/m^2_{W})$ , $x_{t_{\tilde{\chi}}}$ = $(m^2_{t}/m^2_{\tilde{\chi}})$ and 
$x_{\tilde{t}_{\tilde{\chi}}}$ = $(m^2_{\tilde{t}}/m^2_{\tilde{\chi}})$.
Its variation is shown in Figure \ref{fig:susyfig}, as a function of the higgsino (stop-right) mass. It can be noticed that 
$\Delta m_d$ can be large for small values of this mass and values of $tan \beta$ close to unity. 

\begin{figure}
\begin{center}
{\epsfig{figure=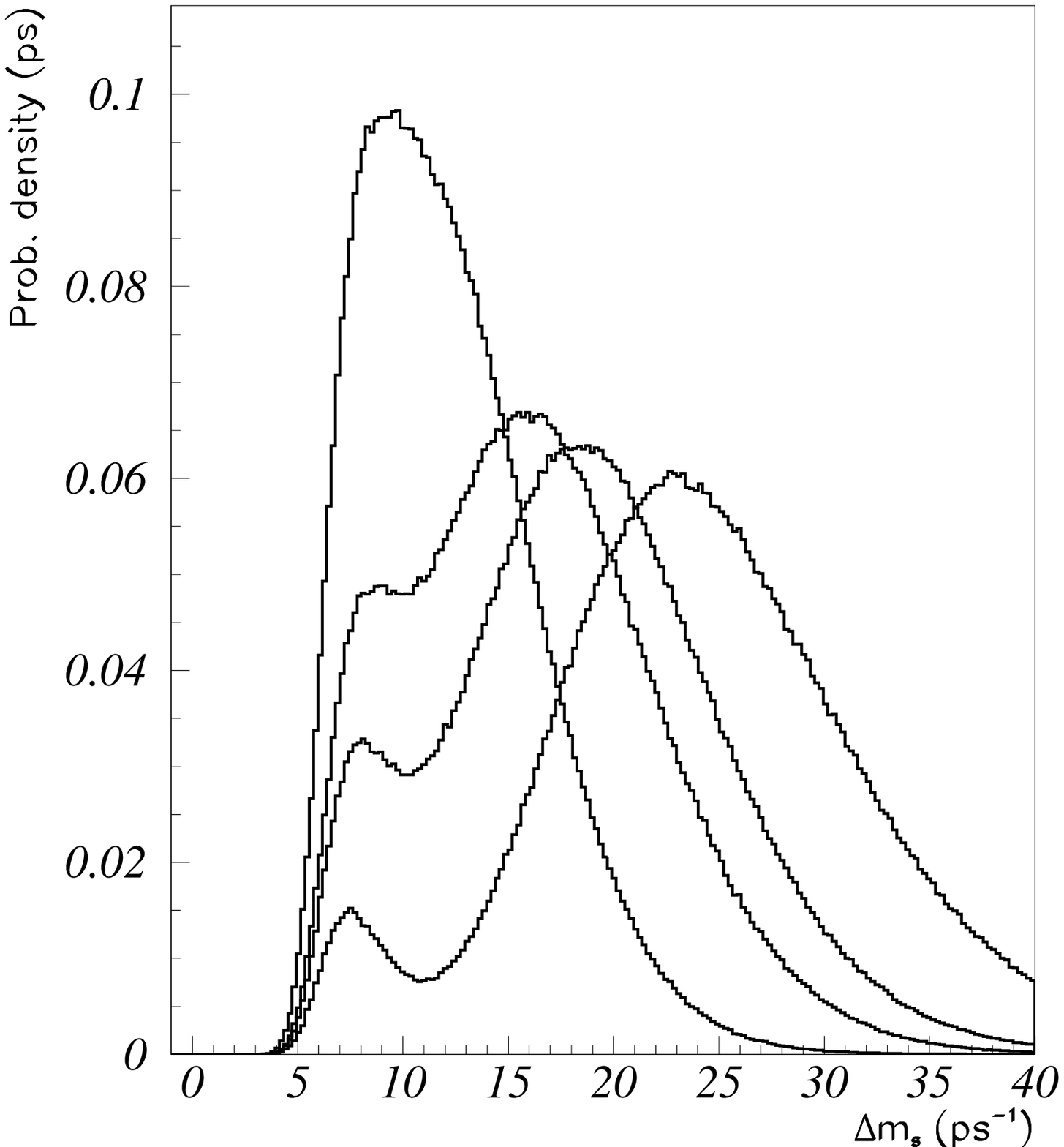,bbllx=12pt,bburx=525pt,bblly=10pt,bbury=520pt,height=8cm}}
\caption{ \it{ $\Delta m_s$ probability distributions obtained with the same constraints as in Figure \ref{fig:hari_1}. The
first distribution, on the left, corresponds to the $\Delta m_s$ probability distribution as obtained in the 
Standard Model ( same as in Figure \ref{fig:deltams} ). The other distributions are obtained in the supersymmetric scenario 
using $tan \beta$ =1.5 and, from left to right, $m_{SUSY}$ = 160, 120 and 80 GeV/c$^2$. }}
\label{fig:dmssusy}
\end{center}
\end{figure}

The present result (\ref{eq:fitsusy}) does not give any stringent constraint on SUSY parameters. 
Sofar no limit on the mass of the lightest supersymmetric particle can be set. This can be clearly seen in Figure \ref{fig:compa}.
The same fit has been repeted for the scenarii presented in the previous section and results have been summarized in Table 
\ref{tab:susy1}. A limit on $m_{SUSY}$ is obtained, using as a central value for $\Delta$, the Standard Model expectation.
Significant limits can be set for small values of $tan \beta$. \\

It is important to stress that also in the SUSY framework the $V_{CKM}$ matrix is unitary. It can happen that the different
constraints, as they are expressed in the Standard Model parametrization,
do not converge at the same point. SUSY can then be invoked. 
Finally it is important to stress that, in this approach, the ratio $\Delta m_d / \Delta m_s$ remains the same as in the Standard 
Model in terms of $\overline{\rho}$ and $\overline{\eta}$. But as, in this scenario, the favoured values for $\overline{\rho}$ and 
$\overline{\eta}$ are not the same as in the Standard Model (see Figure \ref{fig:compa}), 
the value of $\Delta m_s$ will be different. This effect is
shown in Figure \ref{fig:dmssusy}. The $\Delta m_s$ probability distribution moves towards higher values 
for decrasing values of the mass of the lightest supersymmetric particle.

\section{Conclusions.}
\label{sec:9}

The Wolfenstein parametrization describes the $V_{CKM}$ matrix in terms of four parameters :$\lambda$, $A$, $\rho$, $\eta$ and
$\lambda$ is known with a precision better than 1\%. \\
The recent theoretical progress and the new precise measurements of the inclusive lifetime and of the semileptonic branching 
fraction of B hadrons, as well as a new approach in the HQET framework which makes use of the measurement of 
$BR(\mbox{B} \rightarrow \mbox{D}^{*} \ell \nu)$ , allow to get a precise value of $A$: $A$ = 0.81 $\pm$ 0.04.\\
The $\overline{\rho}$ and $\overline{\eta}$ parameters have been determined using the constraints from the measurements of
$  \frac{\left | V_{ub} \right |}{\left | V_{cb} \right |} $, $  \mid \epsilon_K  \mid $ and $\Delta m_d$ and the limit on 
$\Delta m_s$. 
The $\overline{\eta}$ parameter which is related to the CP violating phase is different from zero and is equal to $\overline{\eta} = 0.35 \pm 0.06$.
A further reduction on the uncertainty on $A$ will allow to decrease the uncertainty on $\overline{\eta}$.
The $\overline{\rho}$ parameter is at present poorly known. Its determination has been recently improved by the limit obtained on $\Delta m_s$, the fitted value is $\overline{\rho} = 0.10 ^{+0.14}_{-0.21}$. The uncertainty on  $\overline{\rho}$ is dominated by the precision on the
non-perturbative QCD parameter  $f_B \sqrt{B_{B}}$. It has been shown, that using available constraints from
$  \frac{\left | V_{ub} \right |}{\left | V_{cb} \right |} $, $  \mid \epsilon_K  \mid $, $\Delta m_d$ and $\Delta m_s$, this quantity 
is equal to: $ f_{B_d}\sqrt{B_{B_d}} = 208 ^{+30}_{-40} MeV $. This measurement, which is obtained assuming the validity of the 
Standard Model, improves the theoretical determination of this quantity, $ f_{B_d}\sqrt{B_{B_d}} = 200 \pm 50 MeV$.
This shows that a better determination of $\overline{\rho}$ can be
obtained if  $ f_{B_d}\sqrt{B_{B_d}}$ is measured with a precision better than $\pm 30 MeV$.\\
It has been also shown that the value of $\Delta m_s$ cannot be too large in the Standard Model (Figure \ref{fig:dmssusy}). 
It is expected to lie, within one sigma, in the range between 6.5 and 15 $ps^{-1}$, the most probable value being $ 10~ps^{-1}$.
With the present combined lower limit of $\Delta m_s >~8.0 ps^{-1}$  at 95 $\%$ C.L. LEP experiments are then exploring a very 
interesting region. 
It has been also shown that the introduction of light supersymmetric particles in the loops can significantly 
displace the $\Delta m_s$ distribution towards higher values. \\
A significant comparison between a future direct measurement of $sin 2\beta$ and the present evaluation of this quantity, 
$sin 2\beta~=~ 0.67 ^{+0.12}_{-0.13}$, in the framework of the Standard Model has been also presented. If no progress is obtained
in the evaluation of $f_B$, the evaluation of $sin 2 \alpha$ from the measurement of the sides of the unitary triangle will remain
too uncertain to be compared with the direct measurement.\\
A precise determination of $f_{D_s}$ and $f_{D^+}$ at a future $\tau$-Charm factory can be used to verify lattice QCD evaluations
of these quantities. A precise determination of $f_B$ can then be obtained using these measurements and the extrapolation 
from the D to the B sector, as predicted by lattice QCD. A restricted region can be isolated in the 
($\overline{\rho}$-$\overline{\eta}$) plane which has a similar extension as expected from direct measurement of $sin 2 \alpha$ 
and $sin 2\beta$ with $\pm 0.1$ uncertainty.\\
Finally the present data have been analyzed in a SUSY model, in the framework of a given scenario. Sofar no stringent limits can
be set on the mass of the lightest supersymmetric particles.
The Standard Model fails if the different measurements are not compatible with the same values of $\overline{\rho}$ and 
$\overline{\eta}$. It has been shown that in the SUSY framework different parametrizations 
of  $\mid \epsilon_K  \mid$ and $\Delta m_d$ are obtained providing the extra terms to accomodate this eventual problem.

\subsubsection*{Acknowledgements}

We would like to thank C.W. Bernard for the useful interactions on the subjects related to non perturbative QCD parameters 
and F. Zwirner for clarifications on the theoretical aspects of the paper. We profit all along this work from the
support of F. Richard and D. Treille and we benefit from stimulating discussions with them. They are warmly thanked. 
A special thanks to W. Venus for the useful observations and for the careful reading of the document. 
Finally thanks to the DELPHI Collaboration in which we are working and without which this work could not have been done.


\newpage
 
 
\begin{thebibliography}{ref99} 

\bibitem{ref:sakha}
                  A. D. Sakharov, {\it ZhETF Pis. Red.} {\bf 5} (1967) 32. \\
                  A. D. Sakharov, {\it JETP Lett.     } {\bf 4} (1967) 24. 
\bibitem{ref:universe}
                  P. Huet and E. Sather, {\it Phys. Rev.} {\bf D53} (1996) 4578. \\
                  M. B. Gavela, P. Hernandez, J. Orloff and O. Pene, CERN-TH-94/7368.
%\bibitem{ref:dipole} 
%                  J. Ellis, S. Ferrara and D. V. Nanopoulos, Phys. Lett. B115 231 (1982) \\
%                  S. P. Chia and S. Nandi,    Phys. Lett. B117, 45 (1982) \\
%                  W. Buchmuller and D. Wyler, Phys. Lett. B121, 321 (1983) \\
%                  J. Polchinski and M. B. Wise,      Phys. Lett. B125, 393 (1983) \\
%                  F. del Aguila {\it et al.},        Phys. Lett. B126, 71  (1983) \\
%                  D. V. Nanopoulos and M. Srednicki, Phys. Lett. B128, 61  (1983) 
\bibitem{ref:bjor} B.J. Bjorken and S.L. Glashow, {\it Phys. Lett.} {\bf 11} (1994) 255. \\
                   Y. Hara, {\it Phys. ReV.} {\bf B701} (1964) 134. \\
                   Z. Maki and Y. Ohnuki, {\it Prog. Theor. Phys.} {\bf 32} (1964) 144. 
\bibitem{ref:cabi} N. Cabibbo, {\it Phys. ReV. Lett.} {\bf 10} (1963) 351. 
\bibitem{ref:gim}  S.L Glashow, J.Iliopoulos and L. Maiani, {\it Phys. ReV.} {\bf D2} (1970) 1285. 
\bibitem{ref:jpsi} J.J. Aubert et al., BNL Spectrometer Coll., {\it Phys. ReV. Lett.} {\bf 33} (1974) 1404. \\
                   J.E. Augustin et al., SPEAR Coll., {\it Phys. ReV. Lett.} {\bf 33} (1974) 1406.  
\bibitem{ref:cleo} V. Luth et al.,  SPEAR Coll., {\it Phys. ReV. Lett.} {\bf 34} (1975) 1125. \\
                   G. Goldhaber et al.,  SPEAR Coll., {\it Phys. ReV. Lett.} {\bf 37} (1976) 255.  \\
                   I. Peruzzi et al.,  SPEAR Coll., {\it Phys. ReV. Lett.} {\bf 37} (1976) 569. 
\bibitem{ref:tau}  M. L. Perl et al.,  MARK I  Coll., {\it Phys. ReV. Lett.} {\bf 35} (1975) 1489. \\
                   M. L. Perl et al.,  MARK I  Coll., {\it Phys. Lett. } {\bf 63B} (1976) 466.
\bibitem{ref:fnal}  S.W. Herb et al.,  Fermilab P.C. Coll., {\it Phys. ReV. Lett.} {\bf 39} (1977) 252. \\
                    W. R. Innes et al.,  Fermilab P.C. Coll., {\it Phys. ReV. Lett.} {\bf 39} (1977) 1240. 
\bibitem{ref:bdbd} C. Albjar et al., UA1 Coll., {\it Phys. Lett.} {\bf B186} (1987) 247. \\
                   H. Albrecht et al., ARGUS Coll., {\it Phys. Lett.} {\bf B192 } (1987) 245. \\
                   A. Bean et al., CLEO Coll.,   {\it Phys. ReV. Lett.} {\bf 58} (1987) 183.
\bibitem{ref:top}  F. Abe et al., CDF Coll., {\it Phys. Rev. Lett.} {\bf 74} (1995) 2626.\\
                   S. Abachi et al., D0 Coll., {\it Phys. Rev. Lett.} {\bf 74} (1995) 2632.\\
                   Average by P. Tipton, Plenary Session, ICHEP 96 (25-31 July 1996, Warsaw).
\bibitem{ref:ckm}  M. Kobayashi and T. Maskawa, {\it Prog. Theor. Phys.} {\bf 49} (1973) 652.
\bibitem{ref:pdg}  R.M. Barnett et al. Particle Data Group (PDG) , Review of Particle Properties, {\it Phys. Rev.}
                   {\bf D54}  (1996) 1.
\bibitem{ref:wolf}  L. Wolfenstein, {\it Phys. ReV. Lett.} {\bf 51} (1983) 1945. 
\bibitem{ref:models}  G. Altarelli, N. Cabibbo, G. Corbo, L. Maiani and G. Martinelli, {\it Nucl. Phys.} {\bf B208} (1982) 365.  \\
                      M. Shifman, N.G. Uraltsev and A. Vainshtein, {\it Phys. Rev.} {\bf D51} (1995) 2217. \\
                      M. Luke and M.J. Savage, {\it Phys. Lett.} {\bf B321} (1994) 88. \\
                      P. Ball, M. Beneke and V.M. Braun , CERN-TH/95-65,UM-TH-95-07,HEP-PH/9503492. \\
                      M. Neubert CERN-PPE/96-55.
\bibitem{ref:bigi} I. Bigi, B. Blok, M. Shifman, N.G. Uraltsev and A.I Vainhestein , in " B Decays" edited by S. Stone ,
                   World Scientific (1994) 132 . \\
                   I. Bigi et al., {\it Phys. Lett.} {\bf B323} (1994) 408. \\
                   I. Bigi UND-HEP-95-BIG01. \\
                   M. Neubert, {\it Phys. Lett.} {\bf B338} (1994) 84.\\
                   P. Ball and U. Nierste, {\it Phys. Rev.} {\bf D50} (1994) 5841.
\bibitem{ref:sl1} T. E. Browder and K. Honscheid UH 511-816-95, OHSTPY-HEP-E95-010 (1995).
\bibitem{ref:sl2} J.D. Richman and P.R.Burchat UCSB-HEP-95-08, Stanford-HEP-95-01 (1995). 
\bibitem{ref:lalachi} A. Stocchi LAL 96-38 (May 1996).
\bibitem{ref:varenna} I. Bigi, M. Shifman and N.G. Uraltsev, " Aspects of Heavy Quark Theory", HEP-PH/9703290
\bibitem{ref:life_wkg} B Lifetimes LEP Working Group, Averages for Winter Conferences (March 1997) \\
                      http://wwwcn.cern.ch/$\sim$claires/lepblife.html
\bibitem{ref:ewg}     LEP-ElectroWorking Group CERN-PPE/96-183. 
\bibitem{ref:hqet}    I. Caprini and M.Neubert, {\it Phys. Lett.} {\bf B380} (1996) 376. \\
                      A. Czarnecki, {\it Phys. ReV. Lett.} {\bf 76} (1996) 4124. \\
                      M. Neubert,  {\it Phys. Lett.} {\bf B338} (1994) 84. \\
                      T. Mannel, {\it Phys. ReV} {\bf D50} (1994) 428.\\
                      M. Shifman, N.G. Uraltsev and A. Vainshtein, {\it Phys. Rev.} {\bf D51} (1995) 2217. 
\bibitem{ref:dstar}   H. Albrecht et al.,  ARGUS Coll., {\it Zeit. Phys.} {\bf C57} (1993) 533. \\
                      B. Barish et al.,  CLEO Coll., {\it Phys. ReV.} {\bf D51} (1995) 1014. \\
                      B. Buskulic et al., ALEPH Coll., CERN-PPE/96-150. \\
                      P. Abreu at al,. DELPHI Coll., {\it Zeit. Phys.} {\bf C71} (1996) 531. \\
                      K. Ackerstaff et al., OPAL Coll., CERN-PPE/96-162.
\bibitem{ref:etabar}  A.J. Buras, M.E. Lautenbacher and G. Ostermaier, {\it  Phys. ReV.} {\bf D50} (1994) 3433. \\
                      A.J.Buras , MPI-Pht/95-17 (1995). 
\bibitem{ref:buras0}  G. Buchalla, A.J. Buras and M.E. Lautenbacher MPI-PH/95-104, to appear in ReV. of Mod. Phys.
\bibitem{ref:buras1}  G. Buchalla, A.J. Buras and M.K. Harlander, {\it Nucl. Phys.} {\bf B337} (1990) 313. \\ 
                      W.A. Kaufman, H. Steger and Y.P. Yao, {\it Mod. Phys. Lett.} {\bf A3} (1989) 1479. \\
                      J. M. Flynn, {\it Mod. Phys. Lett.} {\bf A5} (1990) 877.  \\
                      A. Datta, J. Fr\"olich and E.A. Paschos, {\it Z. Phys.} {\bf C46} (1990) 63. \\
                      S. Herrlich and U. Nierste, {\it Nucl. Phys.} {\bf B419} (1994) 292.\\
                      A.J. Buras MPI-PHT/95-88,TUM-T31-97/95 (1995).
\bibitem{ref:buras2}   A.J. Buras, M. Jasmin and P.H. Weisz, {\it Nucl. Phys.} {\bf B347} (1990) 491.
%\bibitem{ref:bk1}      S. Sharpe, {\it Nucl. Phys.} {\bf B34} (1994) 403. \\
%                       N. Ishizuka et al., { \it Phys. ReV. Lett.} {\bf 71} (1993) 24. \\
%\bibitem{ref:bk2}      W.A. Bardeen , A.J. Buras and J.M. G\'erard, {\it Phys. Lett.} {\bf  B211} (1988) 343. \\
%                       J.M. G\'erard , {\it Acta Polonica} {\bf B21} (1990) 257. \\
%                       J. Bijnens and J. Prades , NORDITA-95/11 HEP-PH/95023.
\bibitem{ref:gerard}  C.W. Bernard private communication.
\bibitem{ref:pascal}   P. Paganini PHD Thesis, LAL 96-18 ( May 1996), in french.
\bibitem{ref:fbrev}    C.T. Sachrajda in "B decays" 2$^{nd}$ Edition edited by S. Stone  W.S. (1994) 602. \\ 
                       A. Soni, {\it  Nucl. Phys.} {\bf B47} (1996) 43. 
                       A.J.Buras  MPI-Pht/95-17 (1995). 
\bibitem{ref:fbstu}    A. Duncan et al. FERMILAB-PUB-94/164-T. \\
                       C.W. Bernard,J.N.Labrenz and A.Soni, {\it Phys. ReV.} {\bf D49} (1994) 2536. \\
                       T.Draper and C. Mc Neile , {\it Nucl. Phys.} {\bf B34} (1994) 453.
\bibitem{ref:sumrules} E. Bagan et al., {\it  Phys. Lett.} {\bf B278} (1992) 457. \\
                       M. Neubert, {\it  Phys. ReV.} {\bf D45} (1992) 2451.
\bibitem{ref:relativistic} S. Capstick and S. Godfrey, {\it Phys. ReV.} {\bf D41} (1990) 2856.
\bibitem{ref:qcd_potential} P. Colangelo, G. Nardulli, M. Pietroni, {\it Phys. ReV.} {\bf D43} (1991) 3002.
\bibitem{ref:bbag1} A. Abada et al., {\it Nucl. Phys.} {\bf B376} (1992) 172.
%\bibitem{ref:bbag2} S. Narisson and A. Pivovarov, {\it Phys. Lett} {\bf B327} (1994) 341. 
\bibitem{ref:alibaba}   A. Ali and D. London , DESY 96-140, UdeM-GPP-TH-96-38 (july 1996).
\bibitem{ref:osciwg}  B Oscillation LEP Working Group \\
                      Averages for Hawaii (24-27 March 1997) and Moriond QCD (22-29 March 1997) Conferences 
                      http://www.cern.ch/LEPBOSC. 
\bibitem{ref:amplitude} H.G. Moser and A. Roussarie, {\it Nucl. Instr. and Meth.} {\bf A384} (1997) 491.
\bibitem{ref:bptau_lep} D. Buskulic et al., ALEPH Coll., {\it Phys. Lett.} {\bf B343} (1994) 444. \\
                        (L3 Coll.) L3 note 1988 - Contributed paper to the Warsaw conf.
\bibitem{ref:bptau_cleo} J.P. Alexander et al., CLEO Coll., CLEO-CONF 94-5 submitted to {\it Phys. ReV. Lett.}
\bibitem{ref:dstau}     S. Aoki et al., (WA75 Collaboration), {\it Prog. Theor. Phys.} {\bf 89} (1993) 131.\\
                        J. P. Alexander et al., CLEO Collaboration, CLEO-CONF 95-22. \\
                        J. Z. Bai et al., BES Collaboration, {\it Phys. Rev. Lett.} {\bf 74} (1995) 4599. \\
                        K. Kodama et al., E653 Collaboration, {\it Phys. Lett.}     {\bf B382} (1996) 299.     \\
                        L3 Collaboration, CERN-PPE/96-198 (1996) submitted to {\it Phys. Lett.} {\bf B}. \\
                        Average by J. Richmann, Plenary Session, ICHEP 96 (25-31 July 1996, Warsaw). 
\bibitem{ref:soni} C.W Bernard, J.N. Labrenz and A. Soni, {\it Phys ReV.} {\bf D49} (1994) 2536.
\bibitem{ref:dconstant} C. Alexandrou, S. Gusken, F. Jegerlehner, K. Schilling and R. Sommer, {\it Phys. Lett.} {\bf B256} (1991) 60.\\
                        C. Alexandrou, S. Gusken, F. Jegerlehner, K. Schilling and R. Sommer, {\it Zeit. Phys.} {\bf C62} (1994) 659.\\
                        R.M. Baxter et al., UKQCD Coll., {\it Phys. ReV} {\bf D49} (1994) 1594. \\
                        K.M. Bitar et al., {\it Phys. ReV} {\bf D49} (1994) 3546.\\
                        C. Bernard, T. Draper, G. Hockney and A. Soni, {\it Phys. ReV.} {\bf D38} (1998) 3540.\\
                        M.B. Gavela et al., {\it Nucl. Phys.} {\bf B306} (1988) 677. \\
                        M.B. Gavela et al., {\it Phys. Lett.} {\bf B206} (1988) 113. \\
                        T.A. Grand and D.R. Loft, {\it Phys. ReV.} {\bf D38} (1988) 954. \\
                        A. Abada et al., {\it Nucl. Phys.} {\bf B376} (1992) 172. \\
                        C. Bernard, J. Labrenz and A. Soni, {\it Phys. ReV.} {\bf D49} (1994) 2536. \\
                        H. Hamber, {\it Phys. ReV.} {\bf D39} (1989) 896. \\
                        T. Bhattacharya and R. Gupta (for LANL-Coll.) Proceeding of international conference on lattice field
                        theories "Lattice 94", to be published in {\it Nucl. Phys.} {\bf B} ( Proc. Suppl.).
\bibitem{ref:highl}    M. Canepa et al., DELPHI-Internal Note DELPHI 97-10 PHYS 669.
\bibitem{ref:bigivbc}  I. Bigi and T. Mannel symposium for the 60$^{th}$ Birthday of K. Zalewski to appear in Acta Pysica Polonica.
\bibitem{ref:susy}     A. Brignole, F. Feruglio and F. Zwirner, {\it Zeit. Phys.} {\bf C71} (1996) 679. 



\end{thebibliography}
\end{document}